\documentstyle[psfig]{mn}

\begin{document}

\def\Ha{{H$\alpha$}}
\def\NIIb  {[N${\scriptstyle\rm II}$]$\lambda$6548}
\def\NIIr  {[N${\scriptstyle\rm II}$]$\lambda$6583}
\def\OIIs  {[O${\scriptstyle\rm II}$]$\lambda$3727}
\def\OIIIr {[O${\scriptstyle\rm III}$]$\lambda$5007}
\def\SIIb  {[S${\scriptstyle\rm II}$]$\lambda$6717}
\def\NII   {[N${\scriptstyle\rm II}$]}
\def\OI    {[O${\scriptstyle\rm I}$]}
\def\OII   {[O${\scriptstyle\rm II}$]}
\def\OIII  {[O${\scriptstyle\rm III}$]}
\def\SII   {[S${\scriptstyle\rm II}$]}
\def\SiII  {Si${\scriptstyle\rm II}$}
\def\HeI   {He${\scriptstyle\rm I}$}
\def\HII   {H${\scriptstyle\rm II}$}
\def\HI    {H${\scriptstyle\rm I}$}
\def\CaII  {Ca${\scriptstyle\rm II}$}
\def\MgII  {Mg${\scriptstyle\rm II}$}
\def\CII   {C${\scriptstyle\rm II}$}
\def\CIV   {C${\scriptstyle\rm IV}$}

\def\Hzero  {H$^{\scriptscriptstyle o}$}
\def\Nzero  {N$^{\scriptscriptstyle o}$}
\def\Ozero  {O$^{\scriptscriptstyle o}$}
\def\Hplus  {H$^{\scriptscriptstyle +}$}
\def\Nplus  {N$^{\scriptscriptstyle +}$}
\def\Oplus  {O$^{\scriptscriptstyle +}$}
\def\Hpplus {H$^{\scriptscriptstyle ++}$}
\def\Npplus {N$^{\scriptscriptstyle ++}$}
\def\Opplus {O$^{\scriptscriptstyle ++}$}

\def\deg {$^\circ$}
\def\arcmin {$^\prime$}
\def\arcsec {$^{\prime\prime}$}
\def\spose#1{\hbox to 0pt{#1\hss}}
\def\simlt{\mathrel{\spose{\lower 3pt\hbox{$\mathchar"218$}}
    \raise 2.0pt\hbox{$\mathchar"13C$}}}
\def\simgt{\mathrel{\spose{\lower 3pt\hbox{$\mathchar"218$}}
    \raise 2.0pt\hbox{$\mathchar"13E$}}}
\def\Em{{${\cal E}_m$}}
\def\emunits{{\,cm$^{-6}$\,pc}}
\def\eg{{\rm e.g., }}
\def\aB{{\alpha_{\rm B}}}
\def\Pcos{{$\Phi^\circ$}}
\def\Jcos{{$J_{-21}^\circ$}}
\def\phiI{{$\phi_i$}}
\def\ie{{\rm i.e. }}
\def\etal{{\rm\ et~al. }}
\def\tauLL{{$\tau_{\scriptscriptstyle LL}$}}
\def\Vlsr{{${\rm V_{LSR}}$}}
\def\Vgsr{{${\rm V_{GSR}}$}}
\def\kms{{\,km\,s$^{-1}$}}
\def\Msun{{\,M$_\odot$}}

\def\aj{{AJ}}
\def\araa{{ARA\&A}}
\def\apj{{ApJ}}
\def\apjs{{ApJS}}
\def\apss{{Ap\&SS}}
\def\aap{{A\&A}}
\def\aapr{{A\&A~Rev.}}
\def\aaps{{A\&AS}}
\def\azh{{AZh}}
\def\jrasc{{JRASC}}
\def\mnras{{MNRAS}}
\def\pasa{{PASA}}
\def\pasp{{PASP}}
\def\pasj{{PASJ}}
\def\sovast{{Soviet~Ast.}}
\def\ssr{{Space~Sci.~Rev.}}
\def\zap{{ZAp}}
\def\nat{{Nature}}
\def\aplett{{Astrophys.~Lett.}}
\def\fcp{{Fund.~Cosmic~Phys.}}
\def\memsai{{Mem.~Soc.~Astron.~Italiana}}
\def\nphysa{{Nucl.~Phys.~A}}
\def\physrep{{Phys.~Rep.}}

\title[Optical spectroscopy of an HVC]{The Smith Cloud: HI
associated with the Sgr dwarf?}

\author[J. Bland-Hawthorn \etal]{J. Bland-Hawthorn$^1$, S. Veilleux$^2$,
G.N. Cecil$^3$, M.E. Putman$^4$,\cr
B.K. Gibson$^4$ \& P.R. Maloney$^5$\\
$^1$Anglo-Australian Observatory, P.O. Box 296, Epping, NSW 2121, Australia \\
$^2$Dept. of Astronomy, University of Maryland, College Park, MD 20742, USA \\
$^3$Dept. of Physics \& Astronomy, University of N. Carolina, Chapel Hill, NC 
27599, USA\\
$^4$Mount Stromlo \& Siding Spring Observatories, Weston Creek P.O., Weston, 
ACT 2611, Australia\\
$^5$CASA, University of Colorado, Boulder, CO 80309-0389, USA}
\maketitle
\begin{abstract}
     The Smith high velocity cloud (\Vlsr\ $=$ 98\kms) has been observed
     at two locations in the emission lines \OIIIr, \NIIb\ and
     \Ha. Both the \NII\ and \Ha\ profiles show bright cores due to
     the Reynolds layer, and red wings with emission extending to
     \Vlsr\ $\approx$ 130\kms.  This is the first simultaneous
     detection of two emission lines towards a high velocity cloud,
     allowing us to form the ratio of these line profiles as a
     function of LSR velocity. At both cloud positions, we see a clear
     distinction between emission at the cloud velocity, and the
     Reynolds layer emission (\Vlsr\ $\approx$ 0). The \NII/\Ha\ ratio
     ($\approx$0.25) for the Reynolds layer is typical of the warm
     ionised medium.  At the cloud velocity, this ratio is enhanced by
     a factor of $3-4$ compared to emission at rest with respect to
     the LSR.  A moderately deep upper limit at \OIII\ (0.12R at 
     3$\sigma$) was derived from our data.
     
     If the emission arises from dilute photoionisation from hot young
     stars, the highly enhanced \NII/\Ha\ ratio, the \OIII\ 
     non-detection and weak \Ha\ emission (0.24$-$0.30R) suggest that
     the Smith Cloud is 26$\pm 4$ kpc from the Sun, at a
     Galactocentric radius of 20$\pm 4$ kpc. This value assumes that
     the emission arises from an optically thick slab, with a covering
     fraction of unity as seen by the ionising photons, whose
     orientation is either (a) parallel to the Galactic disk, or (b)
     such as to maximize the received flux from the disk.  The
     estimated mass and size of the cloud are $4\times 10^6$\Msun\ and
     6 kpc. We discuss a possible association with the much larger Sgr
     dwarf, at a galactocentric radius of 16$\pm$2 kpc, which lies
     within 35$^\circ$ ($\sim$12 kpc) of the Smith Cloud.
\end{abstract}

\begin{keywords}
ISM: general --- galaxies: intergalactic medium --- individual object: 
Smith Cloud, Sgr dwarf --- Galaxy: halo --- techniques: interferometric
\end{keywords}

\section{Introduction}

\subsection{Historical context}
Since their discovery thirty years ago (Muller\etal\ 1963), the nature
and origin of high velocity clouds (HVCs) has remained highly
controversial. Their chequered history has been discussed by Verschuur
(1988) and Wakker \& van Woerden (1997).  HVCs $-$ which cover at least
a third of the sky $-$ are concentrations of neutral hydrogen with
velocities which do not conform to a simple model of galactic
rotation.  Few, if any, clouds have reliable distance determinations,
which has encouraged wide ranging speculation as to their origin.
Explanations (assumed distances are given in parentheses)
range from local supernova remnants ($\sim$100 pc),
large-scale expanding motions in nearby spiral arms ($<$ 1 kpc),
condensations in the local galactic halo ($\sim$ 1 kpc), structures in
the galactic warp (5$-$20 kpc), tidal disruptions of the Magellanic
Clouds (20$-$50 kpc), intergalactic gas ($>$ 50 kpc) or protogalaxies
($\sim$500 kpc).

The distance uncertainty continues to be the major stumbling block in
understanding HVCs (Schwarz, Wakker \& van Woerden 1995) since the
cloud density and mass scale inversely with distance, and as the
square of the distance, respectively.  Ferrara \& Field (1994; see
also Wolfire\etal\ 1995) have suggested one possible method for
systematic distance determinations based on the core/envelope
structure observed in some HVCs (Cram \& Giovanelli 1976). Part of the
problem is that, besides \HI\ observations, it has proved difficult to
detect HVCs in other spectral windows (Wakker \& Boulanger 1986;
Colgan, Salpeter \& Terzian 1990).

\subsection{Previous experiments}
Initially, the prospect of optical emission line detections of HVCs
looked bleak.  Reynolds (1987) did not detect any of six
HVCs in the range \Em\ $=0.6-1$ cm$^{-6}$ pc with the Fabry-Perot
`staring' technique. But there have since been many attempts to detect
HVCs at wavelengths other than \HI, with sensitive \Ha\ observations
having the highest success rate.  Wakker \& van Woerden (1997) list
the following published \Ha\ detections of HVCs: HVC 168-43-280 
(0.08 R, Kutyrev \& Reynolds 1989), 
cloud M~II (0.1-0.2 R; Munch \& Pitz 1990), complex C (0.03
R, Songaila\etal\ 1989; 0.09 R, Tufte\etal\ 1998) and the Magellanic
Stream (MS~II, MS~III, and MS~IV, at 0.37, 0.21, and 0.20 R
respectively; Weiner \& Williams 1996).

Other optical detections of HVCs rely primarily on absorption line
studies.  This work can be used to constrain distances
and determine metallicities (i.e. Magellanic
metallicities in Lu\etal\ 1998; see also Schwarz, Wakker \& van
Woerden 1995).  Most detections of visual band and/or ultraviolet
absorption by metal ions are listed in Table 3 of Wakker \& van
Woerden (1997).  The ions most commonly detected are \CaII, \MgII, and
\CIV. van Woerden\etal\ (1997) recently used Ca K absorption to
determine an upper limit on the distance to complex A, giving 4$-$11
kpc as the first distance bracket for an HVC.  HST is frequently used
to search for the UV absorption lines, with the most recent detection
by Sahu \& Blades (1997) of \SiII\ towards HVC 487.

All CO observations of HVCs have resulted in non-detections (Hulsbosch
1978; Giovanelli 1986; Kim\etal\ 1989), indicating that some HVCs are
further than 3 kpc away.  Though previous searches for far infrared
emission have been negative (Wakker \& Boulanger 1986; Bates\etal\ 
1988; Fong\etal\ 1987), Ivesic \& Christodoulou (1997) searched the
IRAS point source catalog and found possible young stellar objects in
the \HI\ cores of the M cloud and complex H.

There have been claims for enhanced x-ray emission towards HVCs from
ROSAT observations, \eg towards Complex M near M~I and M~II
(Herbstmeier 1995), and possibly from Complex C (Kerp 1996).  Blom
(1997) claims extended MeV emission is associated with HVC Complexes M
and A, in the same location as associated diffuse soft x-rays.

Absorption at 21 cm has been more successful for HVC detection. It
yields a measurement of the hydrogen spin temperature, T$_s$, and can
be used to determine the kinetic gas temperature (Dickey 1979; Liszt
1983).  The results of this work generally yield temperatures between
50$-$100K.  Colgan\etal\ (1990) finds T$_s$ $>$ 20-70K for the
Anticentre Clouds, T$_s$ $>$ 20K for HVC 43-13-309, and T$_s$ $>$ 10K
for the Magellanic Stream.  Mebold (1991) find T$_s$ $>$ 50K for
another position in the Magellanic Stream.  Definitive results include
two components at 70K and 300K in Cloud R (Payne\etal\ 1978;
Payne\etal\ 1980) and 50K for complex H (Wakker\etal\ 1991).

We now present the first simultaneous detection of more than one
emission line towards an HVC.  The observations presented here were
first reported by Bland-Hawthorn (1994).  In $\S$2, we
describe the observations before summarising the calibrations and
reductions in $\S$3. In $\S$4, the high velocity cloud measurements
are presented, and these are discussed in the context of ionisation
models in $\S$5.

\section{Observations}

The observations were carried out over three long dark nights (1994
Aug 9$-$12) at the f/8 Cassegrain focus of the AAT 3.9m. Follow-up
observations were carried out at f/15 on 1994 Oct 29-30 and 1995 Sep
27.  The TAURUS-2 interferometer was used in conjunction with three
different Queensgate etalons (Table~\ref{tbl:tbl1}): the HIFI 40$\mu$m
gap etalon from the University of Hawaii, the UMd 44$\mu$m etalon from
the University of Maryland, and the UNC 19$\mu$m etalon from the
University of North Carolina.

\begin{table*}
\caption[TAURUS-2 Observing Log]{\label{tbl:tbl1}
(1) Object name (2) AAT focus (3) exposure times (4) local date of observation
(5) blocking filter / bandwidth (\AA) / tilt angle (6) etalon (7) etalon gap in BCV units (ellipsis
indicates full FSR scanned) (8) average airmass (9) seeing disk FWHM (10) position angle of
detector on the sky.}
\begin{tabular}{lccccccccc}
Object & focus  & exposures    & date   & filter    & etalon & gap & airmass & seeing & P.A. \\
\vspace{0.0mm}\\
\noindent{\it Object fields}\\
Smith \#1 & f/8  & 10 sec       & 9/8/94   & 6555/45/0 & $-$ & $-$   & 1.4   & 1.7     & 0 \\
Smith \#1 & f/8  & 6$\times$20 min  & 9/8/94   & 6555/45/0 & HIFI & 200   & 1.4   & 1.7    & 0 \\
Smith \#2 & f/8  & 10 sec       & 10/8/94  & 6555/45/0 & $-$ & $-$   & 1.4   & 1.3     & 0 \\
Smith \#2 & f/8  & 9$\times$20 min  & 10/8/94  & 6555/45/0 & HIFI & 200   & 1.4   & 1.3    & 0 \\
Smith \#2 & f/8  & 5$\times$20 min  & 11/8/94  & 6555/45/0 & HIFI & 180   & 1.2$-$1.3 & 1.3 & 230 \\
Smith \#2 & f/15 & 20 min       & 29/10/94 & 5012/35/0 & UNC  & 293   & 1.1   & 1.0    & 180 \\
Smith \#2 & f/15 & 20 min       & 30/10/94 & 5012/35/0 & UNC & 300   & 1.15   & 1.1    & 180 \\
Smith \#2 & f/15 & 5 min        & 27/9/95  & 5019/46/0 & $-$  & 260   & 1.2   & 0.9    & 180 \\
Smith \#2 & f/15 & 3$\times$20 min  & 27/9/95  & 5019/46/0 & UNC & 260   & 1.2   & 0.9    & 180 \\
\vspace{0.0mm}\\
\noindent{\it Blank fields}\\
Sky \#1  & f/8  & 6$\times$20 min  & 11/8/94 & 6555/45/0   & HIFI  & 180   & 1.3$-$1.9 & 1.4 & 230 \\
\vspace{0.0mm}\\
\noindent{\it Flux standard: line}\\
N6072    & f/8    & 6 min        & 9/8/94 & 6555/45/0  & HIFI  & $...$ & 1.1   & 1.0    & 0 \\
N6302    & f/8    & 6 min        & 9/8/94 & 6555/45/0  & HIFI & $...$ & 1.1   & 1.3    & 0 \\
N6445    & f/8    & 6 min        & 10/8/94 & 6555/45/0 & HIFI & $...$ & 1.1   & 1.3    & 0 \\
N6563    & f/8    & 6 min        & 10/8/94 & 6555/45/0 & HIFI & $...$ & 1.1   & 1.3    & 0 \\
\vspace{0.0mm}\\
\noindent{\it Flux standard: continuum}\\
LTT 7379 & f/8  & 3$\times$1 min   & 9/8/94 & 6555/45/0 &   HIFI & $...$ & 1.1   & 1.7    & 0 \\
CD-32 9927 & f/8& 3$\times$2 min   & 10/8/94 & 6555/45/0 &  HIFI & $...$ & 1.1   & 1.3    & 0 \\
Feige 110   & f/15 & 2 min     & 30/10/94 & 5012/35/0 & UNC  & $...$  & 1.2   & 2.4    & 180 \\
L870-2      & f/15 & 3 min     & 30/10/94 & 6594/57/0 & UMd  & $...$  & 1.2   & 1.6    & 180 \\
HZ 7     & f/15  & 2 min       & 27/9/95 & 6585/43/8 &  HIFI & 260   & 1.4   & 0.9    & 230 \\
Hiltner 600 & f/15 & 2 min     & 27/9/95 & 6585/43/8 &  HIFI & 260   & 1.3   & 0.9    & 230 \\
Hiltner 600 & f/15 & 2 min     & 27/9/95 & 5019/46/0 &  UNC  & 260   & 1.3   & 0.9    & 230 \\
\\
\end{tabular}
\end{table*}

A single order of interference was isolated using 4-cavity blocking
filters with high throughput (80$-$90\%) and bandpasses well matched
to the etalon free spectral ranges.  The 50mm filters were placed out
of focus close to the focal plane and baffled to give a 5.0\arcmin\ 
field.  The etalon was tilted by 3.4$^\circ$ to shift the optical axis
to the field edge. An in-focus, focal plane, colander mask was used to
ensure that low-order ghosts fall outside the field of view. The
TAURUS-2 f/8 pupil diameter is 59.9mm which is oversized for the
50mm diameter etalon. Due to uncertainties of the precise location of
the optical cavity within the etalon, we placed a 45.0mm aperture stop
immediately in front of the etalon. The pupil stop introduced a major
loss of light (50\%) compounded by a Cassegrain hole which is 17\% of
the total pupil area.  The full pupil was used in the f/15
observations. The observational set-up was the same except for a 75mm
blocking filter centered at 6585\AA\ which was not baffled thereby
producing a field of view similar to the f/8 observations.

Observations were made at two closely spaced positions on the Smith
cloud and at a sky position 15$^\circ$ away (Fig.~\ref{HImom}).  The
etalon was used at fixed gap spacings and the resulting ring pattern
was imaged onto a Tek 1024$^2$ CCD with pixel scales of 0.594\arcsec\ 
pix$^{-1}$ (f/8) and 0.315\arcsec\ pix$^{-1}$ (f/15), with a read
noise of 2.3e$^{\rm -}$. The wavelength range is dispersed quadratically
over the 5\arcmin\ field at 1\AA\ resolution with blue wavelengths to
the north. At both foci, we observed several stellar flux standards
(Table 1); at f/8, we also observed four planetary nebulae.
Observations were also made in the direction of the South Galactic
Pole (Bland-Hawthorn, Freeman \& Quinn 1997, hereafter BFQ), and
twilight flats were taken on all nights with and without the etalon.

\begin{figure*}
\centering
\vspace{15.5cm}
\includegraphics{`zcat}
\caption[]{\HI\ intensity map showing TAURUS-2 fields observed
  towards the Smith cloud where the axes are given in galactic
  coordinates (1950.0). The offset sky position is also shown. The
  black ridge to the north is \HI\ associated with the Galactic plane.
  The column densities at the positions (I,II) are (1.3,1.5) $\times$
  $10^{20}$ cm$^{-2}$.
\label{HImom}}
\end{figure*}

\section{Reduction \& analysis}

\subsection{Spectral extraction}

We begin by establishing the optical axis of the incomplete
calibration rings to better than 0.1 pix.  This was achieved with
orthogonal distance regression.  Next, we azimuthally bin the
calibration rings for all nights to ensure that (a) there were no
opto-mechanical shifts, (b) the instrumental response was constant,
and (c) the etalon gap zero-point was constant from night to night. In
(b), the etalon was found to behave reliably except that there was a
slow drift in the optical gap just after twilight on the second night.
Three exposures for the Smith II field were not used. In total, we
obtained a total of 2.0 hrs on Smith I, 3.7 hrs on Smith II, and 2.0
hrs for the offset sky position.  We also have about six hours of
observation for each night towards the South Galactic Pole (BFQ).

Some calibration spectra showed baseline variations after binning over
different parts of the field. This arises from illumination (vignette)
effects, stray light, chip structure and CCD fringeing which can be
divided out reliably with flatfields.  The spectral bandpass seen by
individual pixels is roughly 1\AA\ where the bandpass centroid
declines by 40\AA\ from the optical axis to the field edge. Twilight
flats were found to be the most reliable except that the Fraunhofer
spectrum leaves its imprint in the data. We divide out the solar
spectrum from the flatfield by establishing the mean spectrum and then
generating a polar image with this spectrum. Occasionally, the
flatfields leave residual fringe structure in the data. It is possible
to remove this with azimuthal smoothing but potentially informative,
intensity variations in the data will be washed out. CCD fringing
constitutes the main systematic error in diffuse detection and
provides a major challenge to achieving deeper detection limits than
that quoted by Bland-Hawthorn\etal\ (1994; hereafter BTVS).  The
response of the blocking filter is removed in the twilight division.

To obtain discrete spectra from the summed CCD images, the data were
divided into annular rings one pixel wide. Within each annulus, we can
calculate the mean, median or mode of the histogram. Cosmic ray events
appear as outliers and are therefore easily removed. We achieved
better results when removing the high density of stars across the
field due to the low galactic latitude. For all fields, we obtained a
short exposure image through the blocking filter without the etalon.
This enabled us to identify compact sources and to mask out these
pixels. Typically 12\% of the pixels were removed in this way.

In order to demonstrate that the flatfielding was adequate, we divided
each summed image into left and right halves and then differenced them
(see Fig.~\ref{left_right}$a$).  The residual for the brightest line
is less than 1\% of peak intensity. Across the full bandpass, the
residual has a dispersion of 0.9 cts. The positive residual feature at
$\lambda$6564.12 appears to be real (see below). In
Fig.~\ref{left_right}$b$, we show our best model fit to the \Ha\
line from the D lamp. Once again, the residual is better than 1\% of
peak intensity.

\begin{figure*}
\centering
\vspace{10.0cm}
\includegraphics{`zcat}
\caption[]{$a.-$ A comparison of the left (top) and
  right (middle) halves of the Smith I field. The difference of these
  two spectra is shown below.  The $+/-$ residual at 6553\AA\ amounts
  to only 1\% of the peak line intensity.  The positive signal at
  6564.12\AA\ (2.4$\sigma$) suggests that the \Ha\ emission in this
  direction has 20\% surface brightness fluctuations on
  arcminute scales.  $b.-$ Fraser-Suzuki (1969) model fit to the \Ha\
  calibration line: the residual is less than 1\% of
  the peak line intensity.
\label{left_right}}
\end{figure*}

\subsection{Self calibration}

While our spectral bandpass is only 40\AA, it comprises typically
3$-$4 OH lines, a geocoronal line, Reynolds layer emission at both
\Ha\ and \NIIb, and 1$-$2 continuum features from scattered solar
and lunar light. These features provide fundamental calibrations of
serendipitous detections within the band.

The OH emission arises from vibrationally and rotationally excited OH
molecules in the mesosphere and lower thermosphere (80-100km altitude)
due to reactions between H$^0$ and O$_3$.  The propagation of density
and temperature differences through the atmosphere can cause large
variations in both the relative and absolute intensity of the OH lines
with periods as short as a few minutes (Adams \& Skrutskie 1996).  In
Fig.~\ref{OH}, we show the temporal dependence of the OH lines, geocoronal
\Ha, and the sky continuum with respect to local midnight for three
consecutive nights.  But the dominant effect in our 20 min exposures
is the increasing average intensity with the zenith angle of the
observation (van Rhijn 1921; Kondratyev 1969). The OH lines in our
bandpass arise from the P$_1$ and P$_2$ branches and vary in an
identical manner.  Osterbrock \& Martel (1992) use OH lines at high
spectral resolution to achieve wavelength calibrations to better than
0.04\AA\, but over our bandpass we achieve 0.01\AA\ at 1$\sigma$ for
the brightest line.

\begin{figure*}
\centering
\vspace{12.0cm}
\includegraphics{`zcat}
\caption[]{Variation in night sky emission towards the
  South Galactic Pole, with respect to local midnight, while tracking
  an object through the meridian. The OH lines follow a secant law,
  the exospheric \Ha\ shows an exponential time dependence, and the
  sky continuum exhibits complex behaviour. Data are shown for three
  consecutive nights in August 1994 (see Table 1): squares (night 1),
  triangles (night 2) and crosses (night 3). In order of decreasing
  intensity, the OH lines for nights 1 and 3 are $\lambda\lambda$6553,
  6577, 6568, and for night 2 are $\lambda\lambda$6553, 6568, 6544.
\label{OH}}
\end{figure*}

There are four discrete components to the \Ha\ region. In Fig.~\ref{OH}, the
geocoronal lines reaches a minimum at local midnight and rises
exponentially towards dawn. Close to twilight or in strong moonlight,
Fraunhofer \Ha\ absorption is clearly seen in the night sky spectrum.
This is only apparent in the first few frames of our data on each
night and is therefore easily corrected for. There are additional 
contributions from the Reynolds layer, and a third redshifted component 
due to the high velocity cloud.

The absolute variation of the geocoronal \Ha\ emission is extremely
well behaved from night to night. We are able to predict the strength
of this atmospheric line to better than 5\% after the first night. The
central wavelength was measured at $\lambda$6562.73 which we could
clearly distinguish from the D lamp wavelength \Ha\ $\lambda$6562.80
at the 3$\sigma$ level (cf. Yelle \& Roesler 1985).  The geocoronal
emission arises from the exosphere at 500 km altitude due to solar
Ly$\beta$ excitation of neutral hydrogen. This process only excites
the 3$^2$P$_{3/2,1/2}$ levels which decay to 2$^2$S$_{1/2}$, in
contrast to the discharge lamp where all fine structure levels are
excited. Since galactic emission is seen in all of our data, it was
necessary to difference an observation near twilight with a later
observation to measure the central wavelength of the exospheric
contribution.

The radial velocity at peak intensity of the Reynolds layer \Ha\ and
\NII\ emission, using new wavelengths from Spyromilio (1995), provides
an independent measurement of the Earth-Sun motion with respect to the
Local Standard of Rest. Indeed, Reynolds (personal communication,
1997) uses this fact to isolate the galactic emission by up to
30\kms\ with a suitable choice of observing season.

Apart from extremely precise wavelength calibration, the geocoronal
and Reynolds layer emission (and to some extent, the sky continuum)
are sensitive to the presence of cirrus. On the third night (Fig.~\ref{OH}),
we stopped observing for a short time an hour after midnight. Note the
presence of cirrus on both the sky level and the geocoronal emission.
Cirrus tends to scatter more light into the beam. For all nights, the
conditions were almost always consistently photometric as judged by
the Reynolds layer emission. The photometric calibration was achieved
using both line and continuum flux standards (Table 1).

For good sky subtraction, we find that it is always necessary to scale
the sky lines by some factor even when interleaving sky with source
observations. (An important exception is telescope nodding coupled
with charge shuffling using an oversized detector now implemented at
the AAT.) So we obtained sequential sky observations at a
comparable airmass after observing the source. After fitting a
Fraser-Suzuki (1969) function to a calibration line, the fit was
used to scale the night sky lines before subtracting from the source
spectra. The Fraser-Suzuki function is a 5-parameter composite of
Lorentzian and Gaussian functions that allows for fine adjustment 
between symmetric and asymmetric line profiles.

The geocoronal \Ha\ uncertainties in subtraction amount to less
than 100 mR at the LSR velocity from frame to frame. The model
subtraction affects the Reynolds emission centroid at the level
of $\sim$1 km/s. This could influence the absolute calibration of 
line fluxes for the Smith Cloud at the level of 20\%. 

BTVS determine the deepest \Ha\ detection possible (10$^{-3}$R at
1$\sigma$)\footnote{1 Rayleigh (1R) is 10$^6/4\pi$ phot cm$^{-2}$
  s$^{-1}$ sr$^{-1}$ or 2.41$\times$10$^{-7}$ erg cm$^{-2}$ s$^{-1}$
  sr$^{-1}$ at \Ha.}  for a diffuse source of emission which fills the
  field. For sources with small projected velocities, this is only
  possible for observations close to midnight and towards the galactic
  poles, in order to suppress both galactic and exospheric emission.
  The current limit is largely set by systematics due to flatfield
  structure.

\section{Results}

\subsection{21cm data}

The Smith Cloud was first identified in 1963 as
having a large positive velocity in the direction of the Galactic
Centre (Smith 1963). 
Wakker \& van Woerden (1991) subsequently noted
that there are smaller \HI\ clouds associated with this complex
(`GCP'), and that the radial velocity is only 25\kms\ in excess of
their galactic rotation model (cf. Fig.~1 of Wakker 1991). 

Integrated HI and individual velocity channel maps for
the Smith Cloud are shown in Figs.~\ref{HImom} and \ref{HIchan}, respectively,
and were constructed from Hartmann \& Burton's (1997)
Leiden/Dwingeloo Survey (LDS).  The cloud has a complex morphology with
the peak in \HI\ intensity at \Vlsr\ $\approx$ 100\kms.  An intense knot is
seen at a galactic latitude $b=13^\circ$ with faint streamers extending
to $b=20^\circ$. The structure is reminiscent of the cometary morphology
of some of the cloud complexes in the Magellanic Stream (Mathewson \&
Ford 1984).

Table \ref{tbl:tbl2} summarises the characteristics of the four primary 
all-sky HVC surveys.
While the LDS may suffer somewhat due to its
higher HI detection limit\footnote{Smoothing the LDS to a velocity resolution
of 16\kms\ would decrease its survey sensitivty to 0.09 K, more in line with
the detection sensitivity of the earlier HVC surveys of B85 and HW88.}, 
its vastly superior velocity resolution makes it 
the only survey suitable for studying the internal dynamics of HVCs.

The PGF98 HVC survey, being carried out in tandem with the HI Parkes All-Sky
Survey (Staveley-Smith 1997), will offer at least an order of magnitude 
improvement over the only previous southern survey (B85), in terms of spatial 
sampling, with the added benefit of a factor of $\sim 5$ increase in survey 
sensitivity and increased resolution.  Compact isolated HVCs, $\sim 80$\% 
of which
would have been missed by B85, will be a unique contribution of the PGF98
survey.  The dramatic improvement in positional accuracy for both existing, 
and newly discovered, southern HVCs will be an important aspect of the PGF98 
work, particularly for future TAURUS-2 emission-line work, as the existing B85
positions are severely compromised by the $2^\circ$ sampling grid.  

In our HVC observing programme, we try to ensure that the small 
field-of-view of TAURUS-2 is actually sampling the peak of the HI emission.
But the new LDS data reveal that we are not always successful.
The TAURUS-2 field positions are indicated in Fig.~\ref{HImom}. The sky
position was chosen to be at roughly the same galactic latitude in
order to match the Reynolds layer contribution. This worked
surprisingly well given that the WHaM survey of the northern sky has
now shown the Reynolds layer emission to be highly structured
(Reynolds\etal\ 1998).

\begin{table*}
\caption[Major surveys for HVCs.]{\label{tbl:tbl2}
Survey references (column 1): B85=Bajaja \etal
(1985); HW88=Hulsbosch \& Wakker (1988); HB97=Hartmann \& Burton (1997);
PGF98=Putman, Gibson \& Freeman (1998).  Columns 2-7 list the telescope 
employed,
its half-power beam-width (HPBW), grid sampling, velocity channel width,
5$\sigma$ brightness temperature detection limit, and 5$\sigma$ column density
detection limit (for an HVC of linewidth 20\kms), respectively.  }
\begin{center}
\begin{tabular}{lllrllc}
\vspace{-0.3mm}
Survey & Telescope     & HPBW               & Grid         & $\;\;\;\Delta v$ & \multicolumn{2}{c}{5$\sigma$ Detection Limit} \\
\vspace{1.5mm}
       &               &                    &              & [\kms] & T$_{\rm B}$ [K] & N$_{\rm H}$ [cm$^{-2}$] \\
B85    & IAR 30m       & $\quad$34$^\prime$ & 120$^\prime$ & $\;\;\;$16 & 0.075 & $2.7\times 10^{18}$ \\
HW88   & Dwingeloo 25m & $\quad$36$^\prime$ & 60$^\prime$  & $\;\;\;$16 & 0.05 & $1.8\times 10^{18}$ \\
HB97   & Dwingeloo 25m & $\quad$36$^\prime$ & 30$^\prime$  & $\;\;\;\;\;$1 & 0.35 & $1.3\times 10^{19}$ \\
PGF98  & Parkes 64m    & $\quad$14$^\prime$ & 6$^\prime$   & $\;\;\;$13.2 & 0.017 & $6.2\times 10^{17}$ \\
\end{tabular}
\end{center}
\end{table*}

\begin{figure*}
\centering
\vspace{15.5cm}
\includegraphics{`zcat}
\caption[]{Velocity channel maps showing kinematic
  structure along the Smith cloud. Most of the \HI\ is confined to the
  LSR velocity range 70 to 115\kms. The greyscale indicates brightness
  temperatures in the range 0 (white) to 5 K (black).
\label{HIchan}}
\end{figure*}

\subsection{Optical data}

In Fig.~\ref{all}, we show the spectra obtained at Smith positions I
and II, and at the offset sky position. Smith II was observed on two
nights, and the spectra are shown separately. The Reynolds layer is
easily detected in \Ha\ at all positions, in \NII\ at Smith I and II,
and only weakly in \NII\ at the offset position. Smith I and the sky
position are affected by Fraunhofer \Ha\ absorption indicated by `F'
on the blue wing. This is easily corrected for due to the large amplitude
change in line flux near twilight.  We superimpose the model for the night
sky emission from scaling our fit to the calibration line. The
residual for each of the four spectra is shown below. We have
superimposed the residual spectra for both nights of the Smith II
observations. These agree remarkably well in accordance with the
excellent photometric conditions.  The residuals at each of the OH
lines amount to 7\% of peak intensity. We have not attempted to
subtract the incomplete OH feature at $\lambda$6577 which falls at the
extreme of the spectral range.

\begin{figure*}
\centering
\vspace{14.0cm}
\includegraphics{`zcat}
\caption[]{$a.-$ Spectrum for the Smith I field overlaid on the modelled
  night sky spectrum. The residual is also shown after subtracting the
  night sky lines.  ``F'' denotes the presence of a Fraunhofer
  absorption feature.  $b.-$ Spectra for the Smith II field. The data
  taken on night 2 have been offset by 30 cts, and the data for night
  3 are shown below. The night sky emission is subtracted as in $a$,
  and the residuals for both nights are plotted together.  These data
  are remarkably consistent given that they were taken on different
  nights, and the geocoronal \Ha\ emission has been subtracted from
  both plots. $c.-$ Spectrum and residual for the sky field.
\label{all}}
\end{figure*}

Unfortunately, a filter was not available at the time of observation for 
the red \NII\ line. However, we achieved an adequate signal to noise ratio 
in the wing of the blue \NII\ line ($3.3\sigma$ per resolution element
on average).  In Figs.~\ref{smith1}-\ref{sky}, we show
magnified regions over the \Ha\ and NII\ lines, and compare to the \HI\
profile as a function of the LSR velocity. The ratio of \NII\ to \Ha\ 
is also shown as a function of LSR velocity. Only a rough comparison can 
be drawn between the optical and \HI\ data. The \HI\ profile is integrated 
over a 30\arcmin\ beam, compared to the 5\arcmin\ beam of the Fabry-Perot 
data.

\begin{figure*}
\centering
\vspace{14.0cm}
\includegraphics{`zcat}
\caption[]{Comparison of \Ha, \NII\ and \HI\ emission with LSR velocity for the
  Smith I field.  (a) \HI\ spectrum for 0.5$^\circ$ field, (b) line
  ratio formed from (c) the \NII\ spectrum, and (d) the \Ha\ spectrum.
  The \NII/\Ha\ ratio (0.25) across the Reynolds Layer is
  characteristic of the warm ionised medium, but the ratio peaks at
  roughly unity close to the \HI\ velocity centroid (90\kms) of the HVC.
  The \Ha\ line has been corrected for Fraunhofer absorption; the effect
  on the line ratios before this correction is shown with open symbols.
\label{smith1}}
\end{figure*}

\begin{figure*}
\centering
\vspace{14.0cm}
\includegraphics{`zcat}
\caption[]{Same as \protect{Fig.~\ref{smith1}} for the Smith II field. 
  The \NII/\Ha\ ratio peaks at 0.7 close to \Vlsr $=$
  75\kms. The observations for the third night are not shown as they are
  entirely consistent with the second night's observations shown here. 
\label{smith2}}
\end{figure*}

At the offset sky position,
no \HI\ emission is seen beyond about 50\kms, and this also holds for the
[NII] and \Ha\ emission.  The [NII]/\Ha\ ratio is roughly 0.15, 
typical of solar-abundance \HII\ regions than the warm ionised medium.
There are several candidates in the NED database which could
conceivably explain this low ratio for the Reynolds layer.

\begin{figure*}
\centering
\vspace{14.0cm}
\includegraphics{`zcat}
\caption[]{Same as \protect{Fig.~\ref{smith1}} for the sky field. The 
  \NII/\Ha\ ratio through the Reynolds layer
  ($\approx$0.15) is more characteristic of solar-abundance \HII\
  regions. The blue wing of the \Ha\ line is shown without the
  Fraunhofer correction.
\label{sky}}
\end{figure*}

In Figs.~\ref{smith1} and \ref{smith2}, extended [NII] and \Ha\ 
emission is seen out to 130\kms\ towards Smith I and II. At Smith I,
the [NII]/\Ha\ ratio reaches $0.9\pm 0.1$ at \Vlsr\ $=$ 100\kms. At
Smith II, [NII]/\Ha\ reaches a somewhat lower value of $0.7\pm 0.1$ at
\Vlsr\ $=$ 80\kms. At both positions, the \HI\ intensity peaks at an
intermediate value of \Vlsr\ $=$ 90\kms. These ratios are factors of
$3-4$ times higher than observed in the Reynolds layer emission
($0.25\pm 0.07$), and $1.5-2.5$ times higher than reported for high
latitude gas in spirals.  This indicates that [NII]$\lambda$6583/\Ha\ 
falls in the range $2.1-2.7$.  Such enhanced `low ionisation' line
strengths for diffuse ionised emission have recently been reported in
the outer parts of spirals (BFQ). In Fig.~\ref{sky}, the [NII]/\Ha\
ratio in the Reynolds emission is a factor of two smaller than the
canonical value. We note that the off-field line-of-sight passes closer
to the Galactic Centre than the Smith Cloud direction, and may therefore
result from direct or scattered emission originating in luminous star 
forming complexes.

In order to determine the [NII] and \Ha\ line fluxes for the Smith
Cloud, we modelled the line ratio dependence with velocity in terms
of two components: a Reynolds layer contribution peaking at $+$8\kms\
with a declining red wing, and a contribution from the HVC peaking
at $+$100\kms\ (Smith I) and $+$80\kms\ (Smith II). 
For the Smith I field, we derive \Em(\Ha) $=$ 0.73\emunits\ (0.24R)
and \Em([NII]) $=$ 0.42\emunits\ (0.14R).  For the Smith II field, we
derive \Em(\Ha) $=$ 0.89\emunits\ (0.30R) and \Em([NII]) $=$
0.52\emunits\ (0.17R).  While the shot noise uncertainty is
approximately 10\%, the formal errors ($25-30$\%) are dominated by the
systematics of our model separation.  The Reynolds layer \Ha\ flux at
both positions is 5.5$\pm$0.1\emunits\ (1.8R).

In Fig.~\ref{o3}, we show the observed spectrum for the Smith II
region in the vicinity of \OIIIr\ (indicated by an arrow).
While we have managed only a moderately deep upper limit at \OIII\
(0.12R at 3$\sigma$), this provides an important constraint on
photoionisation models.

\begin{figure*}
\centering
\vspace{8.0cm}
\includegraphics{`zcat}
\caption[]{Spectrum of Smith II for the \OIIIr\ region.
  No emission was detected at the expected position shown by an arrow.
  The He I $\lambda$5015.68 calibration line is shown in comparison.
\label{o3}}
\end{figure*}

\section{Discussion}

\subsection{Overview}

In this section, we argue that the \Ha\ flux from the Smith Cloud
arises from photoionisation by OB stars dispersed over the Galactic
disk. That the line ratio peaks close to the LSR velocity of the \HI\
peak is strongly suggestive of {\it in situ} photoionisation rather
than shocks. Therefore, the enhanced \NII/\Ha\ line ratio
and non-detection of \OIII\ indicates that the radiation field is very
dilute (cf. Fig. 10 in BFQ). There are at least two interpretations for
this diluteness. First, the Smith Cloud lies close to the Galactic
plane in a region of shadow from strong UV sources.  Secondly, and
more interestingly, the Smith Cloud lies at a large galactocentric
distance. For the latter, the observed \Ha\ emission measure can be
converted to a rough distance once the Galactic halo ionising field is
established (Bland-Hawthorn \& Maloney 1998; hereafter BM).

We have not attempted a Galactic extinction correction for the line fluxes
since the mean column density towards the Smith Cloud (Smith I:
N$_{\rm H} = 1.3 \times 10^{20}$ cm$^{-2}$; Smith II: N$_{\rm H} = 1.5
\times 10^{20}$ cm$^{-2}$) appears to be dominated by the cloud
itself. The TAURUS-2 fields sit on top of the E(B-V)=0.18 contour in
Fig.~5$a$ of Burstein \& Heiles (1982).  While the cloud lies at low
galactic latitude ($b=-15^\circ$), its longitude ($l=+40^\circ$) is
orthogonal to the direction of the warp (see Fig.~\ref{smith_view};
Burton \& te Lintel Hekkert 1986).

\begin{figure*}
\centering
\vspace{17.0cm}
\includegraphics{`zcat}
\caption[]{Location of the Smith Cloud ($b=-15^\circ$) seen from the South 
  Galactic Pole projected onto the Galactic plane.  The circles are drawn
  at radii of 30, 40, 50 and 60 kpc. The galactic
  azimuthal angle $\Theta$ differs from galactic longitude $\ell$ in the
  position of the origin, taken here to be the Galactic Centre.  The
  positions of the Sgr dwarf ($b=-14.5^\circ$) and the Magellanic Clouds
  (LMC: $b=-33^\circ$, SMC: $b=-44^\circ$) are also shown.  The `w'
  labels show where the outer warp in the Galactic \HI\ occurs.  The `?'
  label indicates where a counter feature to the Smith Cloud may fall
  if the latter is associated with the Sgr dwarf.
\label{smith_view}}
\end{figure*}

\subsection{A distance for the Smith Cloud}

We now determine a distance for the Smith Cloud on the basis of the
observed \Ha\ emission. The emission measure ${\cal E}_m$ from the
surface of a cloud embedded in a bath of ionising radiation gives a
direct gauge, independent of distance, of the ambient radiation field
beyond the Lyman continuum (Lyc) edge. This assumes that the covering
fraction {\it seen by the ionising photons} is known (assumed to be
unity) and that there are sufficient gas atoms to soak up the incident
ionising photons. BM present a model for the Galactic halo ionising
field normalized by recent \Ha\ detections along the Magellanic Stream
(Weiner \& Williams 1996).  A summary (and parametric form) of the
model for the Galactic halo is given in the appendix.

Uniformly enhanced \NII/\Ha\ emission in diffuse gas is
generally observed at large distances from the source of ionisation,
in particular, the young stellar disk. Examples include gas in face-on
spirals (\eg Hoopes\etal\ 1996), in edge-on spirals (\eg Veilleux\etal\
1995), and at spiral edges (BFQ).  BFQ (see their
Fig.~10) note that an enhanced \NII/\Ha\ ratio occurs when
the ionisation parameter (${\cal U}=10^4 \varphi_4/c n_{\rm H}$) is
less than 10$^{-4}$. This measures the ratio of the number of ionising
photons $\varphi_4$ (in units of 10$^4$ cm$^{-2}$ s$^{-1}$) to the
number of neutral gas atoms $n_{\rm H}$ (cm$^{-3}$) at the surface of
the \HI\ cloud. The observed \Ha\ emission measures imply $\varphi_4 =
60$ (see appendix) such that we derive a lower limit on the gas density,
$n_{\rm H}$ $\simgt$ 0.2 cm$^{-3}$.  (An upper bound comes from the
\SII\ doublet: when these lines are detected in diffuse emission, they
are always found to be in the low density limit, \ie $\sim10^2$
cm$^{-3}$.) Thus, the diluteness inferred from the line ratio is
consistent with the weakness of the \Ha\ emission, {\it and} with the
non-detection of \OIIIr\ in $\S$4.2.

If the cloud distance is $D_c$, the Galactocentric (cylindrical)
radius is $R_c^\prime$, and the Solar radius is $R_o$ 
(see Fig.~\ref{model}), these
quantities are related via the galactic coordinates, $(\ell,b)$, viz.
\begin{equation}
\label{dist}
D_c= \sec b\ (R_o \cos \ell + \sqrt{R_c^{\prime^2} - R_o^2 \sin^2 \ell})
\end{equation}
for radii $R_c^\prime \geq R_o\sin\ell$. (We reserve $R_c$ for the
true radius to the HVC such that $R_c^2 = R_c^{\prime^2} + Z_c^2$.)
The tangent point ($R_c^\prime = R_o\sin\ell$) in the direction of the
Smith complex is 5.1 kpc from the Galactic Centre and 6.1 kpc from the
Sun.  Fig.~\ref{cloud_dist} illustrates the relation
($D_c/R_c^\prime$) vs.  ($R_c^\prime/R_o$) for different longitudes
($b=0$). For the models below, we adopt $R_o = 8.0$ kpc (Reid 1993).

\begin{figure*}
\centering
\vspace{10.0cm}
\includegraphics{`zcat}
\caption[]{The cloud distance from the Sun, $D_c$, can be related to the
  Galactocentric (cylindrical) radius $R_c^\prime$ and the Solar
  radius $R_o$ through the galactic longitude, $\ell$ (see eqn. 1).
  This relation is illustrated for 5 different values of galactic
  longitude, $\ell=$ 5$^\circ$, 20$^\circ$, 40$^\circ$, 60$^\circ$ and
  80$^\circ$.  The approximate distances of the Smith Cloud and Sgr
  dwarf are shown.
\label{cloud_dist}}
\end{figure*}

The Galactic halo ionisation model is described in detail by BM. The
model comprises an opaque, exponential disk of isotropic UV emitters
encompassed by a hot halo. In computing the expected emission
measures, the main assumption is that there are sufficient gas atoms
to soak up the UV field (N$_{\rm H}$ $\geq$ $3\times 10^{17}$
cm$^{-2}$).  We have idealised the HVC as an \HI\ slab at some angle
to the Galactic plane (Fig.~\ref{model}), where $\pi$ is the angle
between the Galactocentric radius vector {\bf c} and the normal to the
cloud face (Fig.~\ref{model}).  

For the results presented in Fig.~\ref{Em}, we examine two cases: (i)
an \HI\ slab faced downwards lying parallel to the Galactic plane;
(ii) an \HI\ slab whose face is oriented to maximise the received flux
from the disk ($\pi = 0$ in Fig.~\ref{model}). We present results for
two different values of the mean optical depth at the Lyman limit,
\tauLL, measured vertically to the disk: (i) \tauLL = 0, for
illustrative purposes; (ii) \tauLL = 2.8, in order to explain the \Ha\ 
measurements along the Magellanic Stream (BM). The first limit is
ruled out by the neutral gas fraction along the Stream (Bland-Hawthorn
\& Maloney 1997), but we include this model for illustration and
interpolation.

\begin{figure*}
\centering
\vspace{10.0cm}
\includegraphics{`zcat}
\caption[]{A high velocity cloud (HVC) is modelled as an optically thick slab of \HI\ gas 
  inclined at an angle $\pi$ to the radial vector {\bf c} from the Galactic 
  Centre (GC); $\psi$ is the angle
  made by our sight line to the cloud face.  We consider
  the influence of ionising radiation on the upper and lower faces by
  the disk and halo radiation fields. 
\label{model}}
\end{figure*}

\begin{figure*}
\centering
\vspace{12.0cm}
\includegraphics{`zcat}
\caption[]{Predicted emission measure \Em\ as a function of projected
  cloud distance $D_c^\prime$ (the true distance from the Sun to the
  cloud projected in the Galactic plane) in the plane of constant
  galactic longitude, $\ell = 40$\deg. The tangent point to the
  Galactic Centre occurs at $D_c^\prime = R_o\cos\ell$. The models
  include (a) an exponential disk radiation field, and (b) a composite
  halo $+$ disk radiation field.  The solid curves are for an
  optically thick disk (\tauLL = 2.8), the dashed curves for an
  optically thin disk (\tauLL = 0).  From top to bottom (at the
  origin), the curves correspond to constant $Z_c$ values of 5, 10, 15,
  20, 30 and 50 kpc.  The emission measures are for a cloud face
  parallel with the galactic disk.  The thick dot-dash line marked by
  circles relates the observed galactic latitude of the cloud to the
  cloud distance (eqn.~\ref{lat}).  In (c), we show the observed emission
  measures towards an \HI\ slab whose orientation maximises the
  received flux from the disk.  In (d), we show predicted values for
  the upper face of the optically thick slab in Fig.~\ref{model}.  For
  all models, the vertical axes have units of log(cm$^{-6}$ pc); these
  can be converted to log(Rayleighs) by subtracting 0.48.
\label{Em}}
\end{figure*}

In Fig.~\ref{Em}, the predicted emission measures are shown for
different vertical distances, $Z_c$, as a function of $D_c^\prime$,
which is the projection of the true distance from the Sun, $D_c$, in
in the Galactic plane.  The results are corrected for the slab
orientation seen from the Sun's position (Fig.~\ref{model}). The
correction factors are
\begin{eqnarray}
\label{factor}
\sec\psi &=& {{D_c}\over{Z_c}}\ \ \ \ \ \ \ \ \ \ \ \ \ \ \ \ \ \ 
\ \ \ \ \ \ \ \ \ ({\rm downward\ face}) \\
&=& D_c\ \ {\sqrt{R_c^{\prime^2}+Z_c^2}\over{R_c^{\prime^2}+Z_c^2-X_c
    R_o}}\ \ \ \ ({\rm maximising\ face})
\end{eqnarray}
($X_c R_o < R_c^{\prime^2}+Z_c^2$) for which $\psi$ is the angle
between our line of sight and the normal to the cloud face.  The
quantity $X_c$ is the component of {\bf c} in the direction of the
Sun.

The peak in \Em\ offset from the origin occurs at closest approach to
the Galactic Centre ($R_c^\prime = R_o \sin l$).  The expected emission
measures close to the plane fall more rapidly with radius due to the
opacity of the disk. The trend is much slower at larger vertical
distances as the cloud face absorbs flux from most of the disk, and
the secant opacity law has only a small effect.

The galactic latitude of the Smith Cloud ($b=-15^\circ$ $\pm 5^\circ$)
provides a further constraint. From equation (1), the vertical cloud
distance, $Z_c$, is related to the Galactic radius $R_c^\prime$ for a
fixed galactic coordinate position, since
\begin{equation}
\label{lat}
Z_c = D_c \sin b .
\end{equation}
This equation is used to derive the bold curves in Figs.~\ref{Em}$a$
and \ref{Em}$c$. Note that even though the curves go through a
maximum, there is no distance ambiguity for an observed \Ha\ flux due
to the latitude constraint in equation~\ref{lat}.

We evaluate the distance for a mean \Ha\ surface brightness of 0.27R.
The systematic error in separating the HVC from the Reynolds
contribution dominates and leads to a projected distance $D_c^\prime =
25\pm 0.6$ kpc (Fig.~\ref{Em}$a$). This assumes we are observing
through an optically thick facet of the cloud parallel with the
Galactic plane. If instead we are observing the face which maximises
the received flux from the Galaxy (Fig.~\ref{Em}$c$), the derived
(projected) distance is $26\pm 0.5$ kpc. Across both models, and
correcting for the vertical height (equation~\ref{lat}), we determine
that the true cloud distance is $26\pm 1$ kpc.  Prima facie, it may
seem surprising that the maximum flux model does not produce a
significantly larger distance for the cloud. But the correction
factors in equation~\ref{factor} favour the `downward face' model, and
largely compensate for the extra flux received by the `maximising
face' model when $\pi=0$ (see Fig.~\ref{model}).

In practice, we do not doubt that the true uncertainty is considerably
larger.  These models assume that the covering fraction of the HI slab
is unity as seen by the ionising photons, and that the Galactic
dust and ionising sources are smoothly distributed. 
The unobscured UV luminosity from the disk is known
to better than a factor of two (BM). We suspect that the emergent UV
at the distance of the Magellanic Stream is probably good to a factor
of $2-3$. This leads to a much larger uncertainty (Fig.~\ref{Em}) of 4
kpc in our distance estimation. Therefore, we derive a cloud
distance of $26\pm 4$ kpc, which corresponds to a Galactic radius of
$20\pm 4$ kpc (equation~\ref{dist}).

We have looked at the complex problem of cloud geometry and porosity
for a range of cloud distributions
(e.g. fractals; see Pfenniger \& Combes 1994).
The surfaces of these clouds are `heated' by both isotropic and
anisotropic sources of ionisation.  Our simulations produce clouds
that are roughly spheroidal. We find that `volume brightened' clouds at 21 
cm\footnote{We adopt `volume brightened' as the antonym of `limb brightened' 
to indicate that spheroidal clouds appear brighter through the centre due 
to the larger column.}, with large filling factors, can exhibit `limb 
brightening' at \Ha\ over much of the Smith Cloud on arcminute scales. 

The expected contrast level ${\cal C}$ is given roughly by
\begin{equation}
\label{limb}
{\cal C} \sim 2 \left({d{\cal R}}\over{\cal B}\right)^{0.5}
\left({\cal R}\over{\cal B}\right)^{0.5}
\end{equation}
where $d{\cal R}$ is the depth of the ionised layer, ${\cal R}$ is the
radius of curvature at the ionised surface, and ${\cal B}$ is the beam 
size. The plasma thickness is roughly $d{\cal R} \sim (10/n_o)$ pc,
where $n_o$ (cm$^{-3}$) is the neutral gas density at the cloud surface.
For data of sufficient quality,  we recommend dividing up the data to see if 
higher levels of \Ha\ contrast and variance are apparent.

For our data, we require the full 5\arcmin\ field to obtain sufficient
signal. If we take ${\cal R} = 2$\deg\ at the bright tip of the Smith Cloud
(Fig.~\ref{HImom}), equation~\ref{limb} predicts that limb brightening 
should be observable ($n_o \sim 1$).  The observed variance (in
the absence of limb brightening) is expected to be no greater than 50\%
or so. More reliable estimates of HVC distances will come from
complete maps over the projected cloud surface in order to identify
local enhancements due to the surface geometry.

The surface ionisation of clouds close to the disk will be highly
susceptible to the poorly known distribution of dust and UV sources,
not to mention shadowing of the disk by other HVCs. But it
is anticipated that more distant HVCs will reveal more uniform \Ha\
emission over the cloud surface compared to HVCs closer to the disk.
Indeed, the \Ha\ variances, the degree of limb brightening and the
mean flux level, taken together, are needed to improve the reliability 
of the \Ha\ distance method.

For the Smith Cloud flux levels, the Galactic coronal emission
(Fig.~\ref{Em}$b$) is not expected to influence our distance
estimate.  In Fig.~\ref{Em}$d$, we show the expected flux levels on
the upper side of the \HI\ cloud which, for the most part, are
undetectable (BTVS).  Thus we anticipate that, in general, the
observed limb brightening at \Ha\ due to disk ionisation will be
highly asymmetric.

\medskip
Wakker \& van Woerden (1991) determine that the Smith complex extends
over 44 square degrees with a flux density of $2.2\times 10^4$ Jy
\kms. If this cloud lies near the tangent point, the \HI\ mass
is $2.1\times 10^5$\Msun\ (corrected to a Solar radius of 8.0 kpc). For
our revised distance, we derive a cloud mass of $3.6\times 10^6$\Msun.

\subsection{Alternative ionising sources}

A crucial assumption of our model is that the halo radiation field at
the surface of an HVC is dominated by the young stellar disk. But gas
heated by supernovae can puncture the cold disk and escape into the
halo (Norman \& Ikeuchi 1989).  The hot halo gas may form
a galactic wind, a hydrostatic atmosphere around the Galaxy or, more
likely, it may eventually cool, recombine and descend onto the disk,
establishing a galactic fountain (Shapiro \& Field 1976;
Bregman 1980; Houck \& Bregman 1990;
Shapiro \& Benjamin 1991).  In this picture, the Galactic
fountain is largely responsible for the hot corona and may possibly
explain HVCs as material involved in the circulation process (but see
Wakker \& van Woerden 1991).  The cool, descending gas of
the galactic fountain would be detectable at anomalous velocities in
21cm emission.  Theoretically, the maximum velocity of gas streaming
to the disk is roughly 100 km s$^{-1}$.  High apparent velocities are
predicted only for distant clouds, because of the particular mechanism
of acceleration in combination with projection effects induced by
galactic rotation.  Thus, the model can explain the existence of gas
with velocities up to 200 km s$^{-1}$, but not if it is generally
nearby.

The importance of detecting HVCs in more than one emission line cannot
be overstated.  A fountain is expected to produce strong low
ionisation lines (\Ozero,\Oplus,\Nplus, etc.) due to shocks and
perhaps metal-enriched material from the central disk (Shapiro \&
Benjamin 1991). The few HVCs with constrained metallicities have
abundances no more than one third of solar abundance (de Boer \&
Savage 1984; Blades\etal\ 1988).  While metal abundances are
uncertain, the emission lines provide an important diagnostic,
particularly in constraining the ionisation parameter.  Detailed line
ratios have been computed for a range of ionising sources, including
shocks, non-thermal power laws, and hot young stars observed through
an opaque medium (Mathis 1986; Veilleux \& Osterbrock 1987; Sokolowski
1992; Sutherland \& Dopita 1993; Shapiro \& Benjamin 1991). If the
surfaces of HVCs are collisionally ionised, it is unlikely that \Em\ 
can give a useful constraint on distance.

However, as noted in $\S$4.2, the \NII/\Ha\ ratio peaks within 10\kms\ 
of the \HI\ velocity. This is a major problem for shock ionisation
models where the emitting regions arise from downstream gas and are
therefore kinematically separated from the pre-shock material (\eg
Sutherland \& Dopita 1993).  Wolfire\etal\ (1995) discuss alternative
sources of HVC ionisation and heating, but these are not expected to
produce significant optical line emission.

\subsection{Possible association with the Sgr dwarf?}

The Galaxy is encircled by satellite galaxies which appear confined to
one or two great `streams' across the sky (q.v. Lynden-Bell \&
Lynden-Bell 1995).  The most renowned of these are the Magellanic
Clouds and the associated \HI\ stream. All of these are expected to
merge with the Galaxy in the distant future, largely due to dynamical
friction from the extended dark halo (Tremaine 1976). But the most
spectacular accreting satellite, the Sgr dwarf, was discovered only
four years ago by Ibata, Gilmore \& Irwin (1994).

Sgr is approximately 25 kpc from the Sun, and 16$\pm$2 kpc from the
Galactic Centre. The long axis of the prolate body (axis ratios $\sim$
3:1:1) is roughly 10 kpc, oriented perpendicular to the Galactic plane
along $l=6^\circ$, centred at $b=-15^\circ$. Sgr contains a mix of
stellar populations, an extended dark halo (mass $\geq 10^9$\Msun) and
at least four globular clusters (Ibata\etal\ 1997).  In
Fig.~\ref{smith_view}, we illustrate the relative positions of the
Smith Cloud, the Sgr dwarf and the Magellanic Clouds. The Magellanic
Stream runs parallel to $\Theta=90^\circ/270^\circ$ through the
Lagrangian point of the LMC/SMC system. The positions of the other dwarf
spheroidals are given by Majewski (1994). The Smith Cloud and Sgr are
at an identical galactic latitude, and 35\deg\ apart in longitude.

Recent work (Ibata, personal communication) reveals that the angular
size of Sgr is close to 30\deg\ (12 kpc) extending almost
perpendicular to the Galactic plane. The Smith Cloud is about half
this size if the distance from the Sun is comparable; its orientation
is 45\deg\ to the plane.  The projected separation is slightly more
than the projected extent of Sgr. The difference in velocity
($\sim$60\kms) is three times higher than expected for gas in
dynamical equilibrium at 15 kpc around a 10$^9$ solar mass object.  The
total \HI\ mass of the cloud is about 10\% of the visible matter in
Sgr.

If our distance range for the Smith Cloud is reliable, it suggests a
possible association with the Sgr dwarf. In the rest frame of the
Galaxy, for Sgr, \Vgsr\ $\approx$ 172\kms\ (Ibata, Gilmore \& Irwin
1995) compared with \Vgsr\ $=$ 236\kms\ for the Smith Cloud (assuming
a rotation speed of 220\kms\ at the Solar Circle).  The stretched
appearance of the \HI\ is suggestive of a tidal feature in strong
interactions. If the cloud constitutes a gaseous tidal arm, we predict
a counterpart on the same galactic latitude ($b=-15^\circ$) at
longitude $\ell\sim 330^\circ$, and GSR velocity $\sim$ 110\kms. This
putative cloud has an LSR velocity greater than 200\kms\ and would
therefore be labelled a `high velocity cloud'.

A direct association of the \HI\ with Sgr poses problems for our current
understanding of the precursor since the dominant stellar population
is old (Fahlman\etal\ 1996) with [Fe/H] $\approx$ -1 (Whitelock,
Catchpole \& Irwin 1996). Since the Smith Cloud gas falls well beyond
the expected tidal radius of Sgr ($\sim 5$ kpc), an alternative
interpretation is that the Smith Cloud has been dislodged from the
outer \HI\ disk by the impact of the Sgr dwarf.  Such a model may be
consistent with Ibata\etal\ (1997, Fig. 11) who use numerical
simulations to show that Galactic tides can force a dwarf spheroidal
into a prolate shape. Most models favour a short period orbit ($\leq$1
Gyr; Oh, Lin \& Aarseth 1995) implying that Sgr has undergone at least a dozen
perigalacticon passages, consistent with the minimum age of the
dominant stellar population.

If future x-ray surveys detect either emission (\eg Bone, Hartquist \&
Sandford 1983) or absorption (\eg Wang \& Taisheng 1996) towards the
Smith Cloud, this would have important ramifications. Nearby \HI\ clouds
have been observed in shadow against the soft x-ray background (\eg
Burrows \& Mendenhall 1991).  There have even been claims for x-ray
shadows from extragalactic objects (Barber, Roberts \& Warwick 1996).
After repeated attempts, we were not allowed access to the ROSAT high
resolution database in order to check this. The low resolution
(40\arcmin), ROSAT all-sky survey shows no obvious emission or
absorption in the direction of the Smith Cloud.

\section{Conclusions}

The recent detections of the Magellanic Stream in \Ha\ (Weiner \&
Williams 1996) provided Bland-Hawthorn \& Maloney (1998) with the
critical normalisation for the emergent UV flux from the Galactic
disk. This model has been used to predict the ionising field
throughout the Galactic halo. We determine a distance to the Smith
Cloud of 26$\pm$4 kpc on the basis of the \Ha\ flux alone. In support
of this picture, the \NII/\Ha\ is greatly enhanced at the velocity of
the cloud and \OIII\ must be very weak, which together indicate a
dilute ionising field impinging the cloud surface.
 
Possible complications are cloud geometry, porosity, and uncertain
extinction corrections. A lower mean disk opacity would put the 
clouds further away. We anticipate that our model is more reliable for
clouds at greater vertical distances as this tends to average out
structure in the distribution of UV sources. Clouds within a few kpc
of the Galactic plane could be in relative shadow, particularly for
opaque disk models.

More accurate distances for HVCs will come from large \HI\ surveys
with optical follow-up. A crucial development has been the
availability of target lists with higher sensitivity and better
resolution, in particular, the Parkes \HI\ Multi-beam Survey
(Staveley-Smith 1997). More recent detections at the AAT are
sufficiently strong to suggest that a significant number of clouds
should be observable with this technique.  As such, it provides a
crucial test of the Blitz-Spergel model (Blitz\etal\ 1996;
Spergel\etal\ 1996) which places roughly half of all high velocity
clouds at extragalactic distances ($\sim 1$ Mpc) within the Local
Group, in which case none of the clouds should be detectable at these
levels.

We have set out to present a simplified picture as a challenge to
theorists and experimentalists alike.  The observational programmes that
we have described here only require small to medium-sized telescopes
(0.5$-$4m) since both HVC and Stream clouds subtend large angles, and
the line surface brightness is expected to be fairly constant over
large angular scales ($\leq 1^\circ$). Since the Fabry-Perot
interferogram is binned azimuthally in order to produce the detection,
the site does not require good seeing although it should have
relatively good photometric stability.  A dry site is favoured because
variable water vapour features can complicate sky subtraction for \Em $<$ 1
cm$^{-6}$ pc. We encourage a more widespread interest in the
Fabry-Perot `staring' technique as it is set to make a profound
contribution to the understanding of both galactic and extragalactic
radiation fields over the coming years.

\section*{ACKNOWLEDGEMENTS}
We are indebted to G. da Costa for clarifying recent developments in
the study of dwarf spheroidals.  We thank R. Ibata, B. Koribalski and 
C.G. Tinney for their assistance with important references, and a
highly competent referee for recommendations which encouraged us to be
more circumspect in our analysis of errors.

\appendix
\noindent\section{Ionisation models}

We present a useful approximation to the Galactic halo ionising field
deduced by BM for an opaque, exponential disk of isotropic UV emitters. 
The disk has a scale length of 3.5 kpc and a radial extent of 25 kpc. 
The dust is assumed to be distributed smoothly throughout the 
exponential disk.  The hot gas distribution is described by a non-singular
isothermal sphere with a central gas density of $2\times 10^{-3}$ cm$^{-3}$,
a scale radius 10 kpc and a halo temperature of 0.2 keV. This density
and temperature distribution matches the emission measure and x-ray
luminosity of the diffuse component determined by Wang \& McCray (1993).
The corona is expected to be negligible for $R^\prime_c < 30$ kpc
although this depends on the Galactic latitude of the cloud. 

In order to derive \Ha\ emission measures, we assume an electron
temperature T$_e \approx 10^4$K, as expected for gas photoionised by
stellar sources, for which the Case B hydrogen recombination
coefficient is $\aB \approx 2.6 \times 10^{-13} (10^4/T_e)^{0.75}$
cm$^3$ s$^{-1}$. At these temperatures, collisional ionisation
processes are negligible. In this case, the column recombination rate
in equilibrium must equal the normally incident ionising photon flux,
$\aB n_e N_{H^+} = \varphi_i$, where \phiI\ is the rate at which Lyc
photons arrive at the cloud surface (photons cm$^{-2}$ s$^{-1}$),
$n_e$ is the electron density and $N_{H^+}$ is the column density of
ionised hydrogen.

The emission measure is ${\cal E}_m = \int n_e n_{H^+}\;dl =n_e
n_{H^+} L\ {\rm cm^{-6}\; pc}$ where $L$ is the thickness of the
ionised region.  The resulting emission measure for an ionising flux
\phiI\ is then ${\cal E}_m = 1.25\times 10^{-2} \varphi_4 \ {\rm
  cm^{-6}\; pc}$ ($= 4.5 \varphi_4 \ {\rm mR}$) where $\varphi_i =
10^4 \varphi_4$.  For an optically thin cloud in an isotropic
radiation field, the solid angle from which radiation is received is
$\Omega = 4\pi$, while for one-sided illumination, $\Omega=2\pi$.  For
our disk model, however, $J_\nu$ is anisotropic and $\Omega$ can be
considerably less than $2\pi$.

For the metagalactic UV field, \Jcos\ is the ionising flux density of the
cosmic background at the Lyman limit in units of 10$^{-21}$ erg
cm$^{-2}$ s$^{-1}$ Hz$^{-1}$ sr$^{-1}$; \Pcos\ ($=\pi$\Jcos$/h$) is
the equivalent photon flux at face of a uniform, optically thick slab.

To a good approximation, the poloidal UV radiation field from the disk is
\begin{equation}
\varphi_4 = 2.8\times 10^6 \ e^{-\tau}\ {R_{\rm kpc}}^{-2} \cos^{0.6\tau+0.5} \Theta \ \ \ \ \ \ \ \ {\rm phot\ cm^{-2}\ s^{-1}}
\end{equation}
where $\Theta$ is the polar angle ($0 \leq \Theta < \frac{\pi}{2}$),
$R_{\rm kpc}$ is the radius in kiloparsecs, and the 
Lyman limit
optical depth $\tau \leq 10$. It follows that the solid-angle averaged
flux is
\begin{equation}
\bar{\varphi_4} = {{2.8\times 10^6\ e^{-\tau}}\over{(0.6\tau+1.5) {R_{\rm kpc}}^2}} \ \ \ \ \ \ \ \ {\rm phot\ cm^{-2}\ s^{-1}}.
\end{equation}
In order to explain Magellanic Stream \Ha\ detections, our model favours 
$\tau \approx 2.8$.

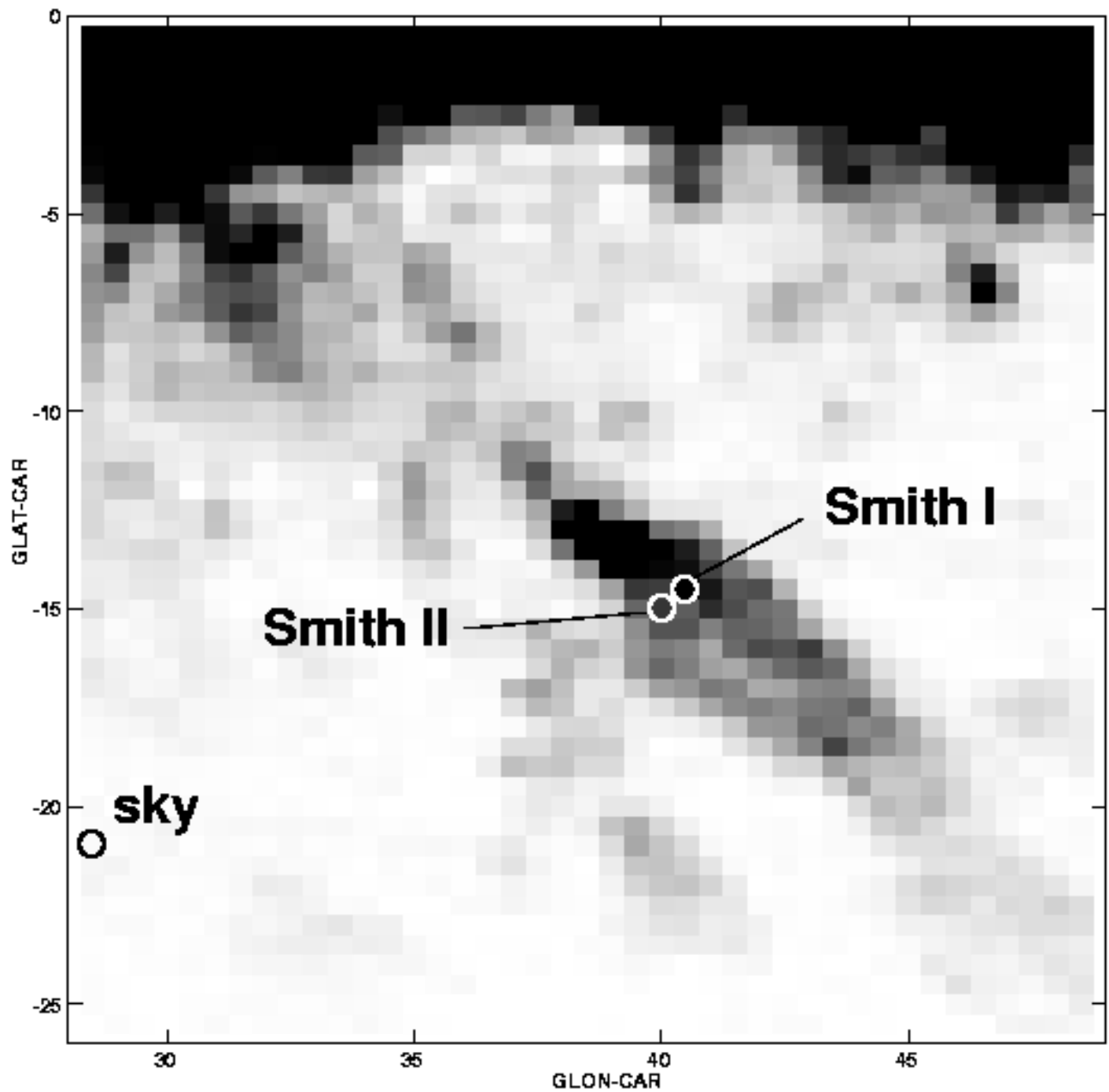
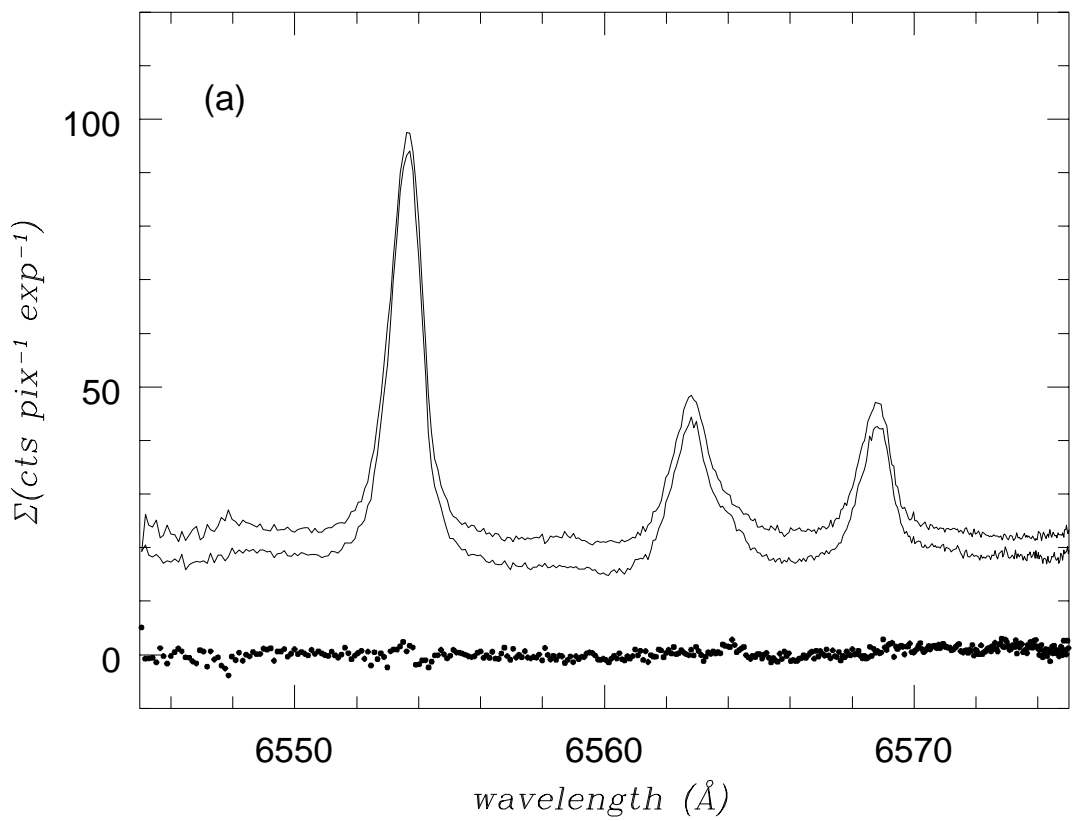
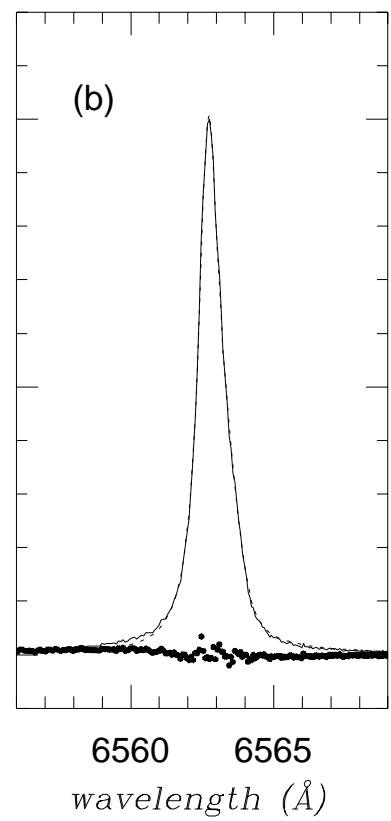
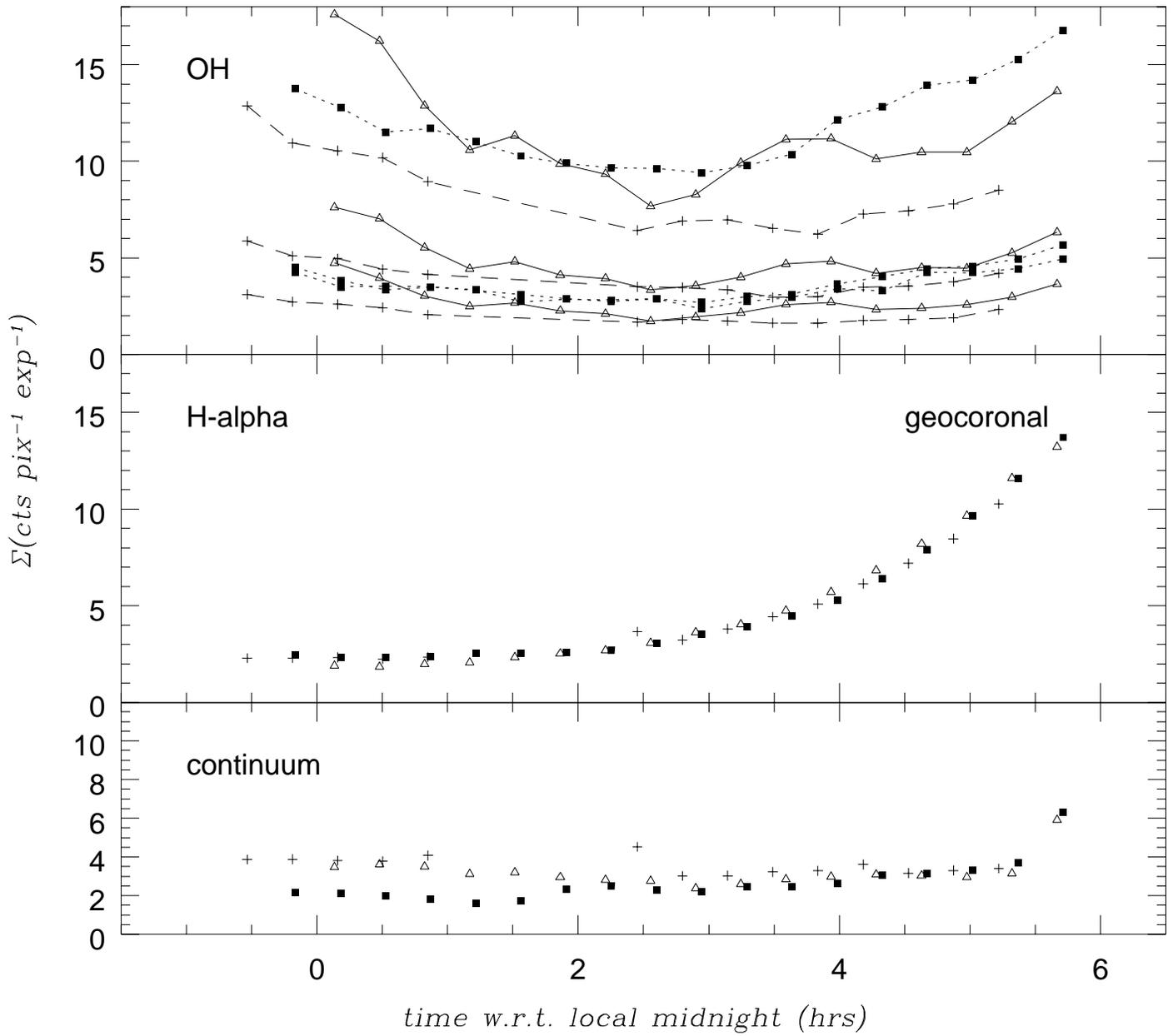
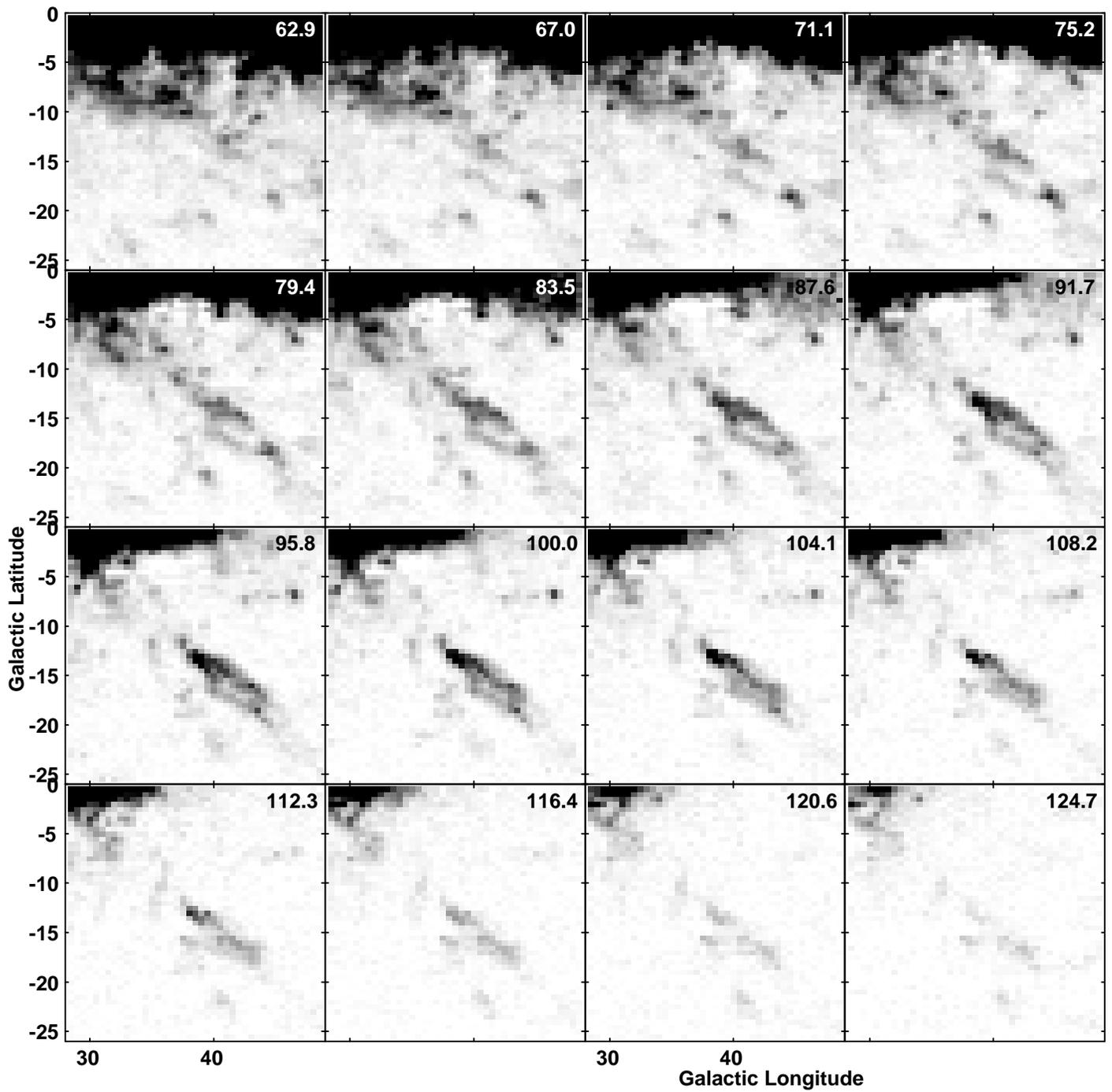
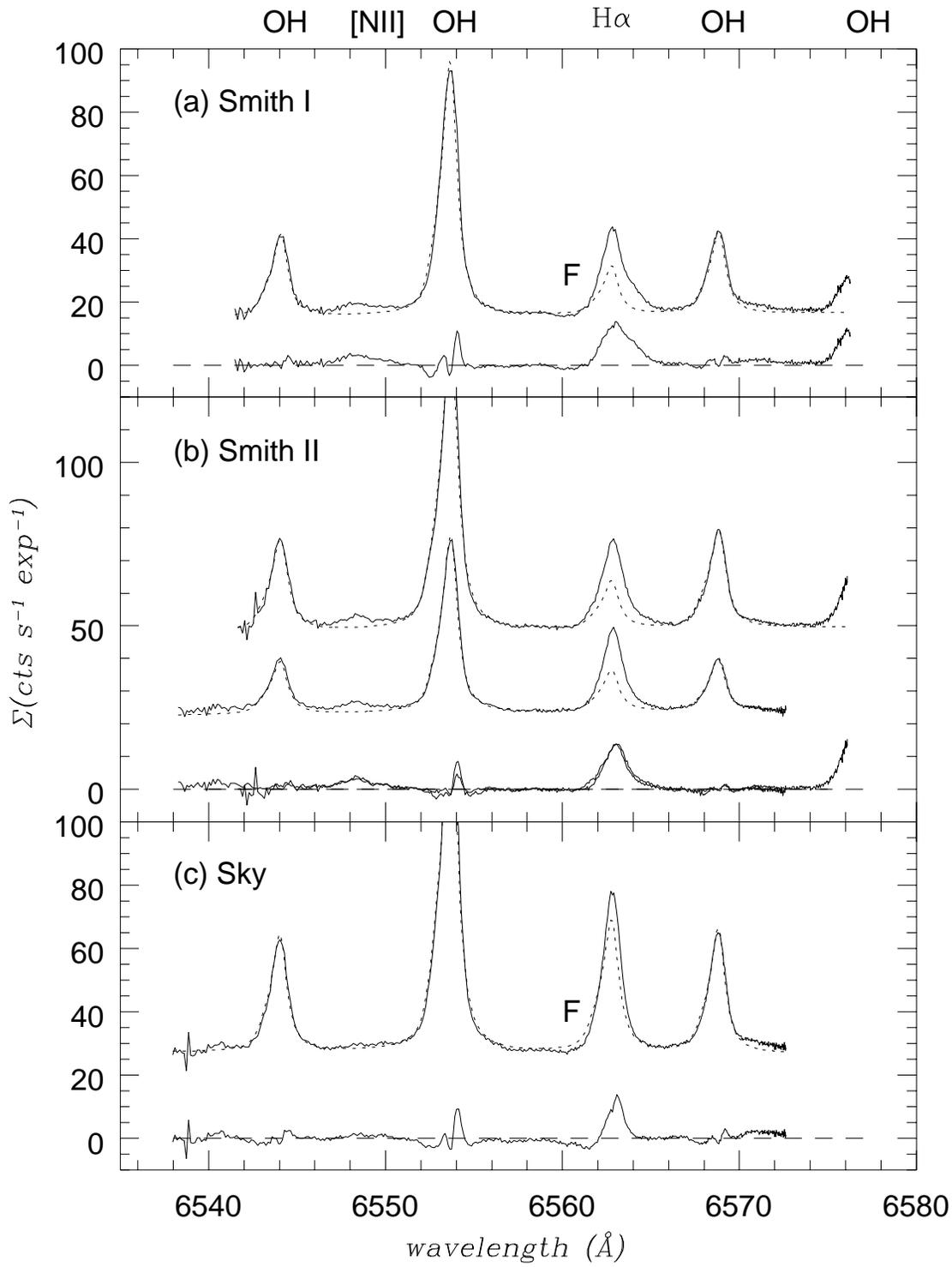
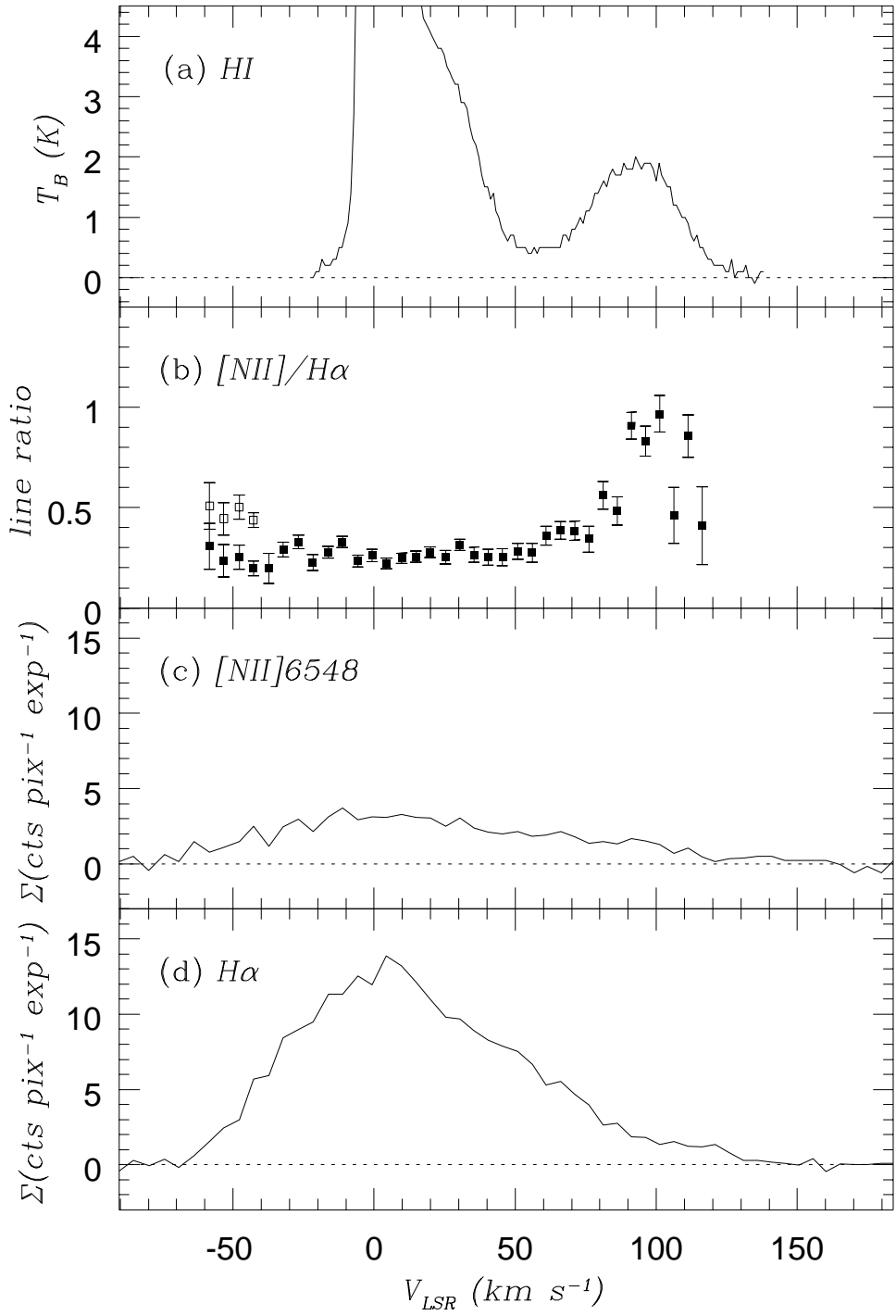
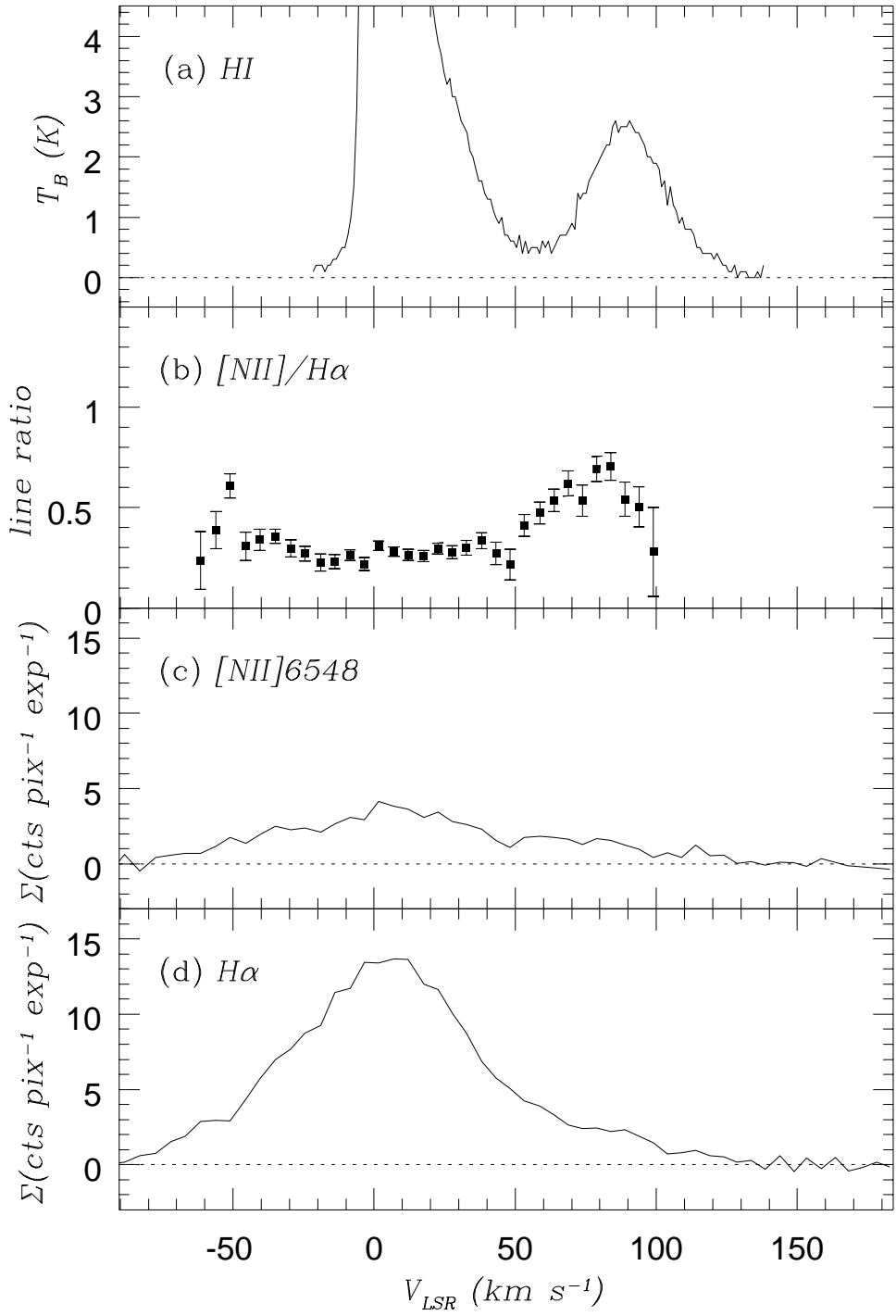
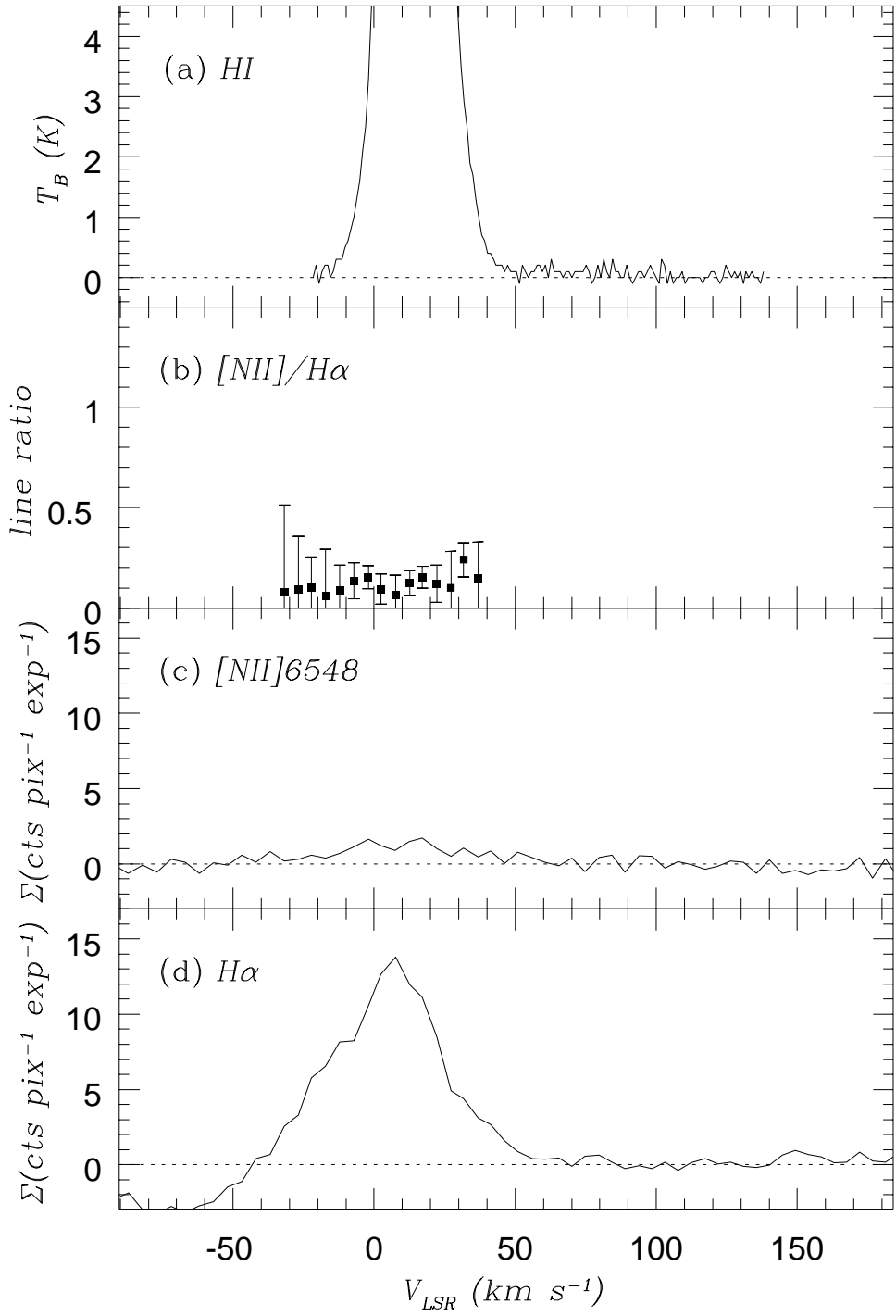
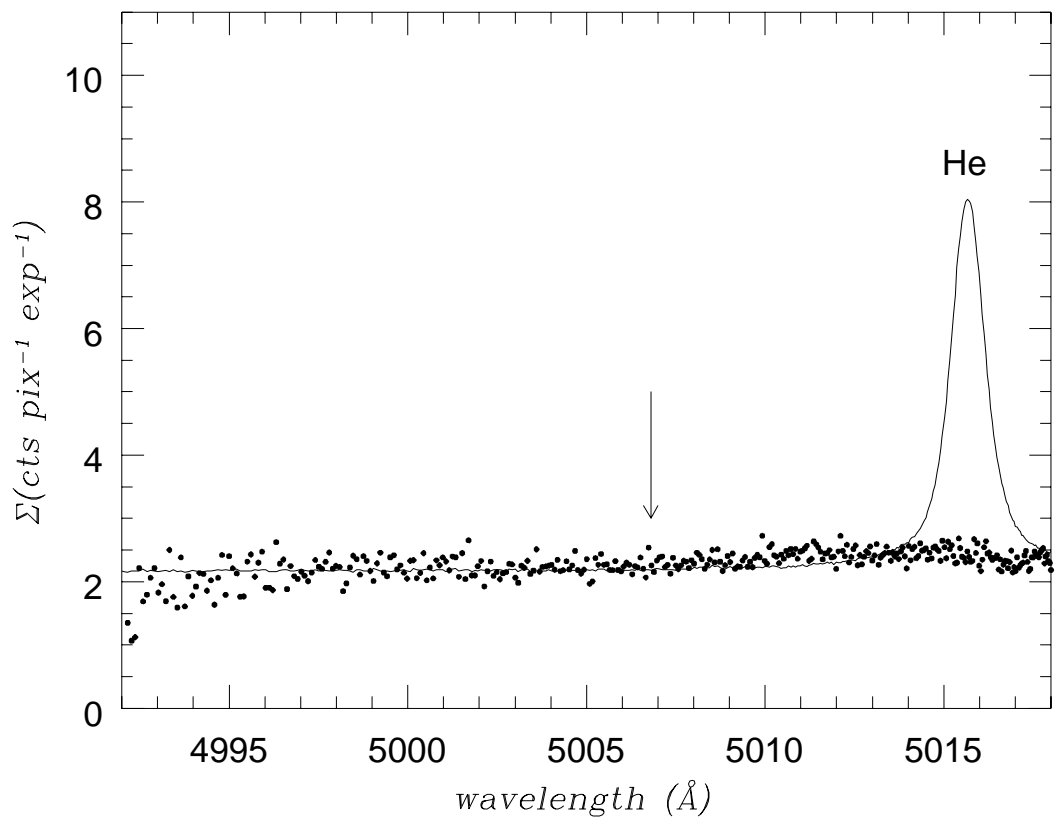
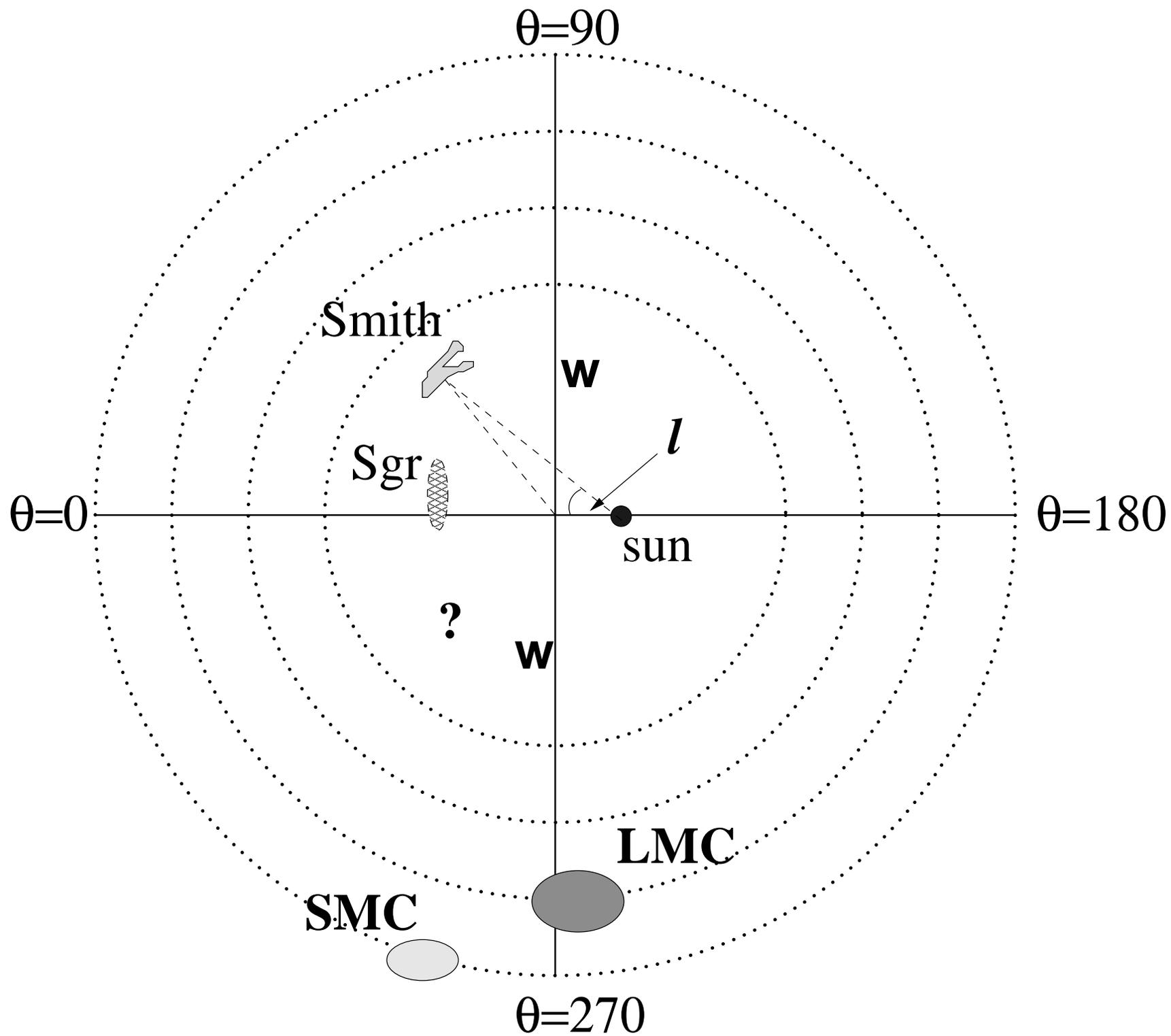
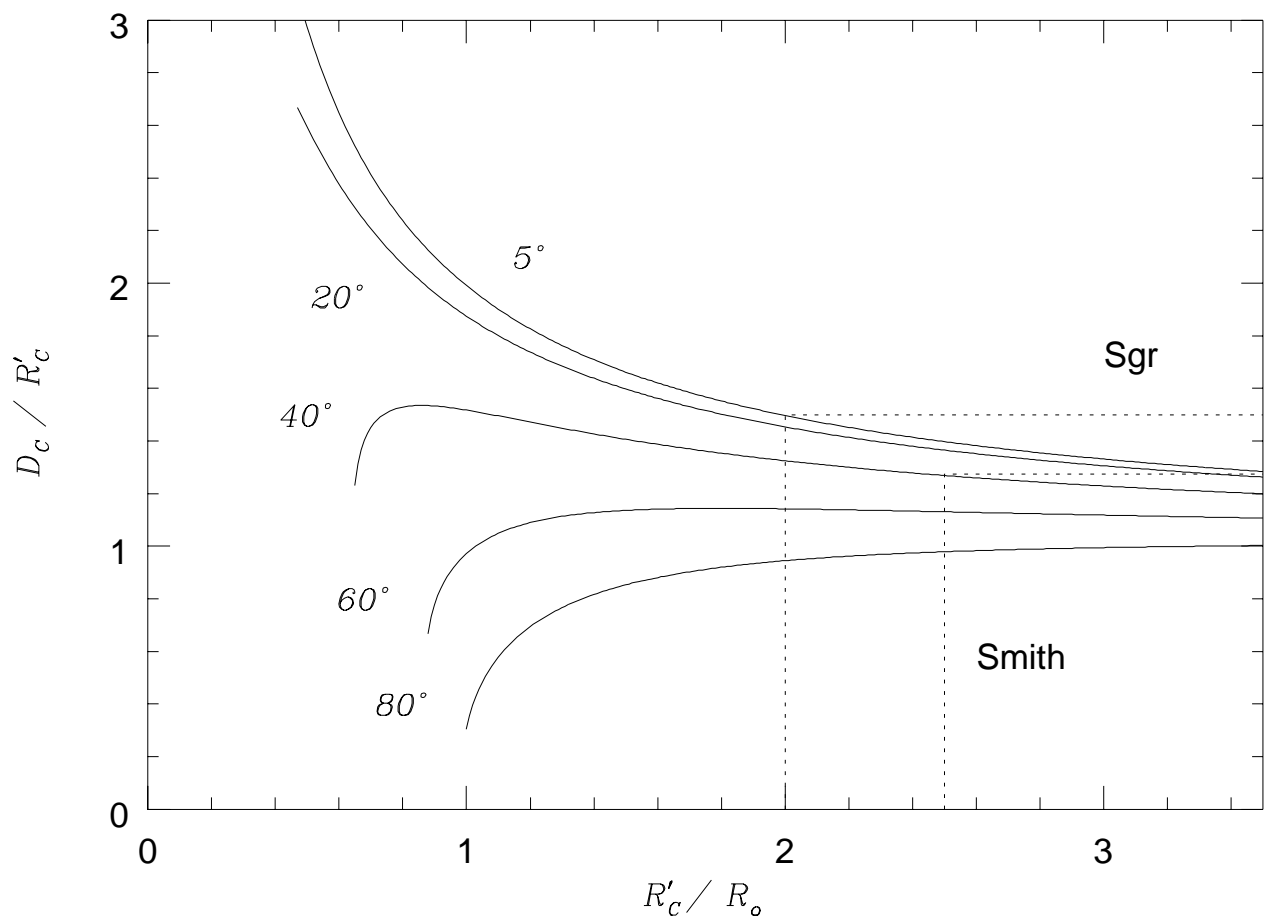
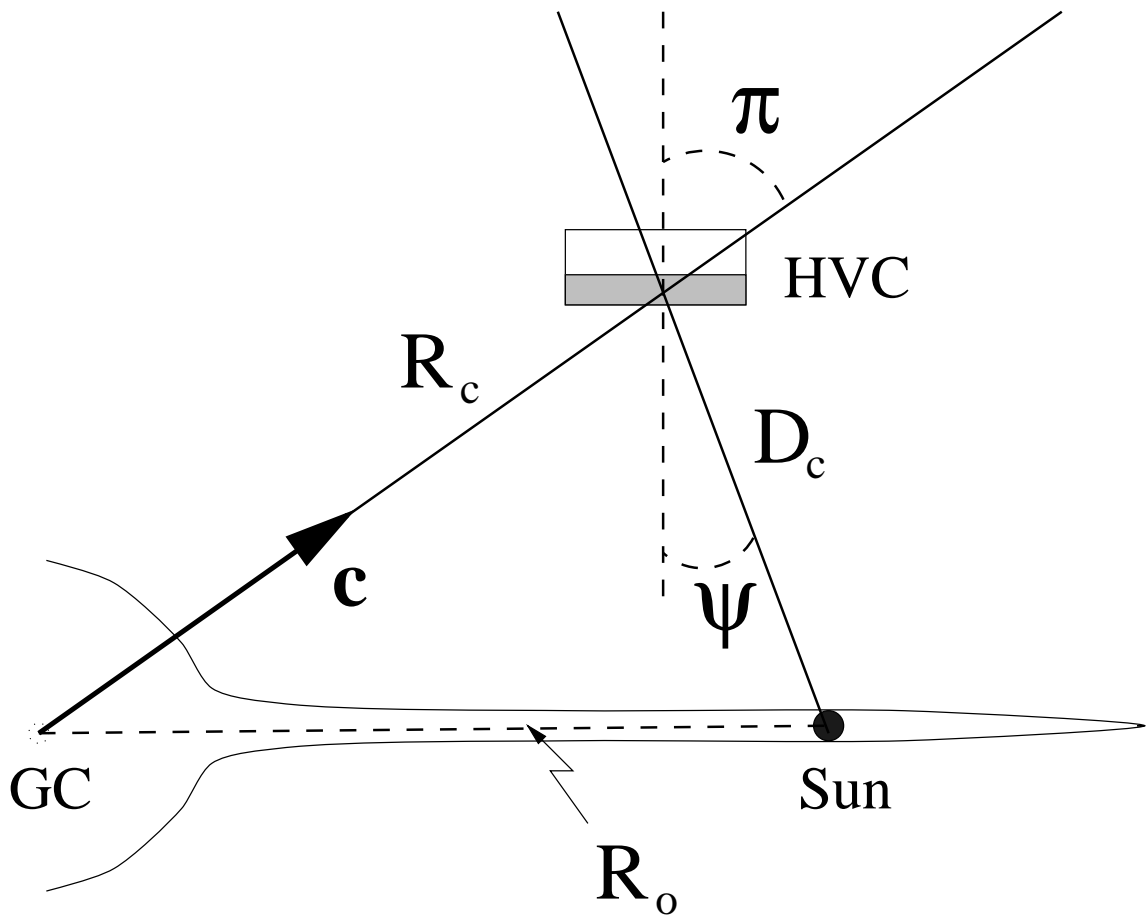
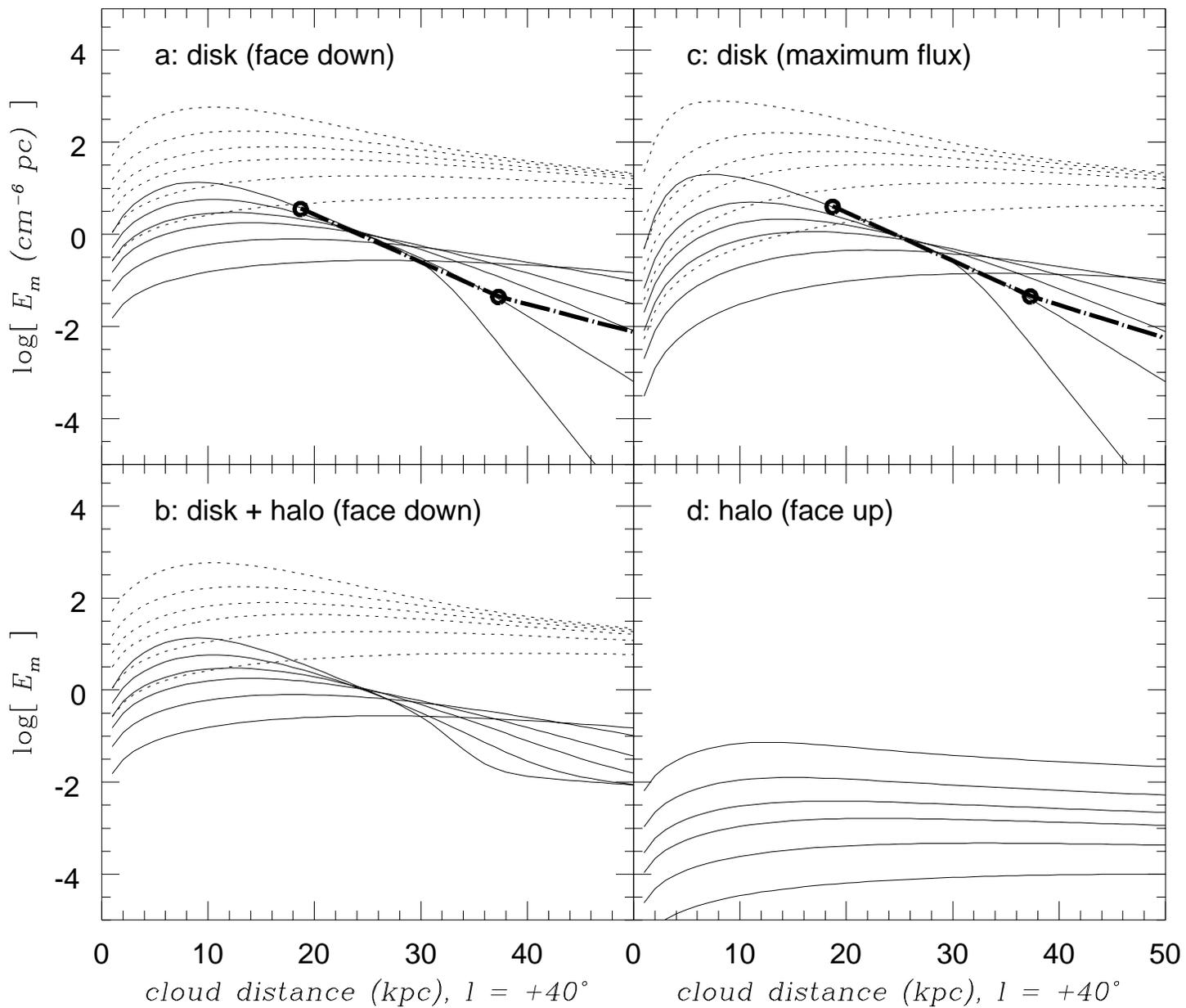
\end{document}